\documentclass[%
 reprint,
 amsmath,amssymb,
 aps,
pra,
]{revtex4-2}

\usepackage{graphicx}
\usepackage{dcolumn}
\usepackage{bm}

\usepackage{xcolor}
\usepackage{xspace}
\usepackage{braket}
\usepackage{upgreek}

\newcommand{\tramp}{t_{\text{r}}}		
\newcommand{\dis}[2]{\epsilon^{#1}_{#2}}	
\newcommand{\pd}[1]{\partial_{#1}}	
\newcommand{\pdx}{\partial_{\bm{x}}}
\newcommand{\pdtildex}{\partial_{\tilde{\bm{x}}}}
\newcommand{\pp}{\partial}
\newcommand{\dd}{\mathrm{d}}
\newcommand{\landau}{\mathcal{O}}
\newcommand{\tr}{\mathrm{Tr}}
\newcommand{\trans}{^\mathrm{T}}
\newcommand{\e}{\mathrm{e}}
\newcommand{\I}{\mathrm{i}}
\newcommand{\re}{\mathrm{Re}}
\newcommand{\im}{\mathrm{Im}}

\newcommand{\ini}{^\text{ini}}
\newcommand{\fin}{^\text{fin}}
\newcommand{\plexp}{\theta}

\newcommand{\drho}{\delta\rho}
\newcommand{\drhotilde}{\delta\tilde\rho}
\newcommand{\vecdrho}{\bm{\drho}}
\newcommand{\Pitot}{\Delta S}
\newcommand{\tsc}{\tau}
\newcommand{\taugamma}{\tau}
\newcommand{\vecj}{\bm{j}}
\newcommand{\vecJ}{\bm{J}}
\newcommand{\veczeta}{\bm{\zeta}}
\newcommand{\vecq}{\bm{q}}

\newcommand{\vecx}{\bm{x}}
\newcommand{\diffunit}{\mathcal{D}}

\newcommand{\integers}{\mathbb{Z}}    

\newcommand{\mathperiod}{\,.}
\newcommand{\mathcomma}{\,,}

\newcommand{\appref}[1]{App.~\ref{#1}}
\newcommand{\figref}[1]{Fig.~\ref{#1}}
\newcommand{\secref}[1]{Sec.~\ref{#1}}

\makeatletter

\newcommand\thefontsize[1]{{#1 The current font size is: \f@size pt\par}}

\makeatother

\begin{document}


\title{Entropy production for quasi-adiabatic parameter changes\\dominated by hydrodynamics}

\author{Philipp S. Wei{\ss}}

\author{Dennis Hardt}
\email{hardt@thp.uni-koeln.de}
 
\author{Achim Rosch}
\affiliation{%
Institut f\"{u}r Theoretische Physik, Universit\"{a}t zu K\"{o}ln, Z\"{u}lpicher Stra{\ss}e 77, 50937 K\"{o}ln, Germany
}

\date{\today}

\begin{abstract}
A typical strategy of realizing an adiabatic change of a many-particle system is to vary
parameters very slowly on a time scale $\tramp$ much larger than intrinsic equilibration time scales. In the ideal case
of adiabatic state preparation, $\tramp \to \infty$, the entropy production vanishes.
In systems with conservation laws, the approach to the adiabatic limit is hampered by hydrodynamic long-time tails, arising from the algebraically slow relaxation of hydrodynamic fluctuations.
We argue that the entropy production $\Delta S$ of a diffusive system at finite temperature in one or two dimensions is governed by hydrodynamic modes resulting in $\Delta S \sim 1/\sqrt{\tramp}$ in $d=1$ and $\Delta S \sim \ln(\tramp)/\tramp$ in $d=2$. In higher dimensions, entropy production is instead dominated by other high-energy modes with $\Delta S \sim 1/\tramp$.
In order to verify the analytic prediction, we simulate the non-equilibrium dynamics of a classical two-component gas with point-like particles in one spatial dimension and examine the total entropy production as a function of $\tramp$.
%
\end{abstract}

\maketitle



\section{\label{sec:intro} Introduction}
The concept of adiabaticity plays a major role in physics ranging from the classification of quantum states of matter 
to the foundation principles of thermodynamics.
For example, the  Carnot cycle describing an idealized heat engine 
is based on a sequence of isothermal
and adiabatic processes.
It is also of practical importance for many experiments ranging from specific heat measurements
to the preparation of correlated states in ultracold atom experiments.

Two types of adiabatic processes are usually discussed.
In quantum mechanics, one considers an isolated system prepared in the ground state $\ket{\psi_0(\lambda)}$ of a Hamiltonian $H(\lambda)$
and a slow change of the Hamiltonian parameter $\lambda \to \lambda^\prime$
within the ramp time $\tramp$.
The adiabatic theorem \cite{Born1928,Kato1950,Avron1999} states that
the initial ground state $\ket{\psi_0(\lambda)}$
transforms into the new ground state $\ket{\psi_0(\lambda^\prime)}$ if $\lambda$ is changed on a time scale much slower
than $1/\Delta$, where $\Delta$ is the (finite-size) gap to excited states of the system.

In the field of thermodynamics, adiabatic processes are intimately linked to the concept of entropy \cite{Callen1985,Lieb1999}.
For an isolated system, an adiabatic process connecting two states $X \to X^\prime$ is characterized by the conservation of entropy, $\Delta S = S(X^\prime)-S(X)=0$. 
In contrast to the quantum adiabatic theorem, one considers classical and quantum mechanical interacting many-particle systems at finite temperature
or finite energy density. Entropy production vanishes in the limit $\tramp \to \infty$. The goal of our study is to investigate $\Delta S$ for large values of $\tramp$ by studying time scales much larger than the time scale $\tau$ for local equilibration, $\tramp\gg \tau$.


Quasi-adiabatic protocols approaching the adiabatic limit are of practical interest for experiments in which the Hamiltonian parameters can be controlled such as ultracold atoms on optical lattices \cite{Bernier2009,Sorensen2010, Lubasch2011,Chiu2018}.
Usually, the goal is to prepare a desired state with low entropy.
Some correlated states (e.g. an antiferromagnetic phase)
are not easily produced by direct loading of atoms into an optical lattice.
An alternative approach is to first generate an easily producable state and then change the Hamiltonian very slowly until the desired state is reached.
However,
the state preparation cannot be performed arbitrarily slowly since the system cannot be isolated for arbitrarily long times.
The optimal ramp time $\tramp$ is a compromise between minimizing the internal entropy production ($\tramp \to \infty$) and minimizing other perturbations like the increase of entropy due to external heating or -- in ultracold atom systems -- atomic losses which become minimal for $\tramp \to 0$.
The question arises at which rate the adiabatic limit is reached if $\tramp$ is increased more and more. What is the asymptotic behavior of $\Pitot(\tramp)$ as a function of $\tramp$?

For quantum systems with a gapped spectrum at zero temperature
the answer can be given in the spirit of the adiabatic theorem.
If the ground state is protected by the gap $\Delta$,
the probability to create excitations can be exponentially suppressed by smooth parameter changes following the Landau-Zehner paradigm \cite{Landau1932,Zener1932}.
Therefore, we expect $\Pitot(\tramp) \sim \e^{-\Delta \tramp}$ such that the adiabatic limit is reached for long enough ramp times $\tramp \gg \Delta^{-1}$.
However, the quantum adiabatic theorem 
does not imply an exponential suppression of the entropy production in gapless many-body systems
\cite{Avron1999,Altland2008,Polkovnikov2008,Moeckel2010,Eckstein2010}.
Entropy production is not exponentially suppressed if a many-body system at $T=0$ is gapless.
Then the entropy production is only weakly suppressed following a power law $\Pitot(\tramp) \propto \tramp^{-\plexp}$, $\plexp >0$ \cite{Polkovnikov2008,Eckstein2010}.
For small parameter changes the exponent $\plexp$ can be calculated perturbatively \cite{Eckstein2010}.
For smooth ramps, $\plexp$ only depends on the low-energy spectrum of the initial state.
If the ramp shape is not sufficiently smooth,
high-energy modes predominantly contribute to the entropy production even for long ramp times $\tramp \to \infty$.
In this case $\plexp$ is fully determined by the ramp spectrum.

A more dramatic situation can occur for non-interacting bosonic systems as has been pointed out by Polkovnikov and Gritsev  \cite{Polkovnikov2008}.
They could identify situations where the number of exciations created during a slow ramp and, therefore, also the entropy diverges in the thermodynamic limit, $\Pitot(\tramp) \propto L^{\delta+1} \tramp^{-\plexp}$ with $\delta>0$.
In this situation the thermodynamic limit $L\to \infty$ and the adiabatic limit $\tramp \to \infty$ do not commute  suggesting that adiabatic processes with zero entropy production cannot be realized for arbitrarily slow parameter changes. The authors also showed numerically that some of these effects
even occur in a Bose Hubbard model \cite{Polkovnikov2008} if the interactions are increased linearly in time starting from a non-interacting model.

Another widely studied problem are slow parameter changes across classical or quantum phase transitions, where the critical slowing down associated with a phase transition and the Kibble-Zurek mechanism \cite{Dziarmaga2010,Deffner2017} leads to non-analytic corrections and an entropy production
again characterized by power-laws, $\Pitot(\tramp) \propto \tramp^{-\plexp}$, where $\plexp$ depends on the universality class and critical exponents of the phase transition.

We will focus on a much simpler situation and consider the case of finite temperature omitting all complications arising from phase transitions or Goldstone modes. If one assumes that {\em all} modes relax to equilibrium with a finite rate $1/\tsc$,
then deviations from equilibrium due to slow changes of parameters are proportional to $\tsc$ and the rate by which changes occur scales with $1/\tramp$. The rate of entropy production
is therefore proportional to $1/\tramp^2$ which results in a total change of entropy proportional to  $1/\tramp $ (see \appref{app:entr-prod-Boltz} for a calculation in the context of the Boltzmann equation). This estimate is, however, based on the
assumption of exponential relaxation of all modes which is rarely met: systems with conservation laws (e.g., energy or particle number conservation) always possess hydrodynamic modes with arbitrarily small relaxation rates. 

Hydrodynamic modes, i.e., diffusive modes and sound waves, are arguably the most common and most generic source of non-analytic
behavior as they occur in any many-body system with local conservation laws. For such systems one can use hydrodynamic equations to describe all properties on long time and long length scales as long as the system remains sufficiently close to thermal equilibrium. (Here, we assume that the temperature $T$ is finite.)
An important consequence of hydrodynamics is the presence of so-called hydrodynamic long-time tails \cite{Alder1967,Alder1970,Ernst1971,Ernst1976,Ernst1976a,Pomeau1975,Forster1975,Harrison1986}. These long-time tails describe, for example, the slow decay of certain correlation functions. More importantly, hydrodynamic modes
provide a bottleneck for equilibration. After a quench, i.e. a sudden change of parameters, a generic interacting many-body system does not
relax exponentially towards thermal equilibrium but approaches thermal equilibrium only algebraically. In a diffusive, $d$-dimensional system, for example, typical observables approach their thermal expectation value only slowly, $\sim t^{-d/2}$ \cite{Lux2014}. These hydrodynamic long-time tails arise from the long-distance fluctuations of hydrodynamic modes.
Thus, it takes a long time to build up the local thermal fluctuations of conserved quantities. 

As hydrodynamic long-time tails dominate equilibration in the long-time limit, one can also expect that they can give a dominant contribution
to the entropy production when parameters of a many-body system are changed slowly. In the following, we will consider a simple hydrodynamic system
which contains only diffusive modes related to energy- and particle number conservation. We compute analytically the contribution to the entropy production due to the hydrodynamic modes and compare our results to numerical simulations of a one-dimensional (1D) classical model.

\section{\label{sec:hyd} Hydrodynamic description of quasi-adiabatic parameter changes}

\subsection{\label{sec:hyd_eq} Hydrodynamic equations}

We consider a closed diffusive system in $d$ dimensions at finite temperature $T>0$ where the parameters of the underlying Hamiltonian change smoothly on a time scale $\tramp \gg \tsc$, much larger than the relaxation time $\tsc$ in which a local equilibrium state is established. Under this condition, the system always stays close to equilibrium and hydrodynamic approaches apply. Our goal is to calculate corrections to the entropy production arising in a translationally invariant system where no macroscopic currents are generated. Thermal and non-thermal fluctuations do, however, activate hydrodynamic modes.

The starting point of our analysis are the continuity equations for energy density $e$ and a finite set of additional densities $n_a$, $a \in \{A,B,...\}$, that remain conserved during the parameter change: 
\begin{eqnarray}
\pd{t}\delta e + \pdx \cdot \vecj_e
&=& 
\delta r_e
\mathcomma \label{eq:cont-e}
\\
\pd{t}\delta n_a + \pdx \cdot  \vecj_a
&=&
0
\mathperiod \label{eq:cont-AB}
\end{eqnarray}
$\delta e = e -\braket{e}$ and $\delta n_a = n_a -\braket{n_a}$ denote the fluctuations of the densities around their average values $\braket{e}=E/L^d$, $\braket{n_a}=N_a/L^d$ where $L$ is the linear system size. $\vecj_e$ and $\vecj_a$ are the corresponding current densities.
The average energy density changes in time according to
$\pd{t}\braket{e} = \braket{r_e}$, where $r_e$ simply parametrizes the rate of change of the energy density. Energy is not conserved as the underlying Hamiltonian is time dependent. For the same reason also $\delta e$ is not a conserved density and we therefore have to include the source term $\delta r_e$ in Eq.~\eqref{eq:cont-e} (to be specified below). $\braket{r_e}$ and $\delta r_e$ are only finite during the time when the parameters of the Hamilton are changed, $0 \leq t \leq \tramp$, but vanish for $t>\tramp$.
In contrast, $N_a=\int \! \dd^d x \,n_a $ is always exactly conserved and one obtains the standard continuity equation for this quantity.

Hydrodynamics is a theory valid on time and length scales large compared to the microscopic scattering times and the microscopic mean-free path. Here we consider inelastic scattering processes arising from interactions, which relax the system towards a thermal equilibrium state.
As a consequence of the coarse-grained description,
the current densities are fluctuating quantities that can be decomposed into two contributions:
\begin{subequations}
\begin{eqnarray}
\vecj_e
&=& 
\vecJ_e[\pdx e,\pdx n_a,\pdx e^2,...]
+
\veczeta_e
\mathcomma \label{eq:current-e}
\\
\vecj_a
&=& 
\vecJ_a[\pdx n_a,\pdx e,\pdx n_a^2,...]
+
\veczeta_a
\mathperiod\label{eq:current-n}
\end{eqnarray}
\end{subequations}
The constitutive relations $\vecJ_e$, $\vecJ_a$ state
that the \emph{average} current densities are caused by inhomogeneous distributions of energy and the conserved quantities $N_a$.
In the long-wavelength limit of a diffusive system, the linear terms are the most relevant contributions of the gradient expansion
\begin{align}
\vecJ_i = -\sum_j D_{ij}(t) \pdx \rho_j + \landau(\pdx\rho^2)\mathcomma
\end{align}
with $\rho_i \in  \{ e , n_a \}$. The diffusion constants (and similar quantities discussed below)
explicitly depend on time as the underlying Hamiltonian is time dependent. For $\tramp \gg \tsc$ and to leading order in $1/\tramp$, they can, however, be obtained from the properties
of the time-independent Hamiltonian $H(t_0)$ with $t_0=t$.

In addition, random transitions of particles between adjacent volume cells give rise to the noise terms $\veczeta_e$, $\veczeta_a$ with $\braket{\veczeta_e} =\braket{\veczeta_a}=0$.
As noise is caused by fast modes on the short time scale $\tsc \ll \tramp$,
the noise correlations can -- to leading order in $1/\tramp$ -- be calculated again from $H(t_0)$ using the fluctuation-dissipation theorem
\begin{eqnarray}
&&
\braket{\zeta_i^{l}(\vecx,t) \zeta_j^{l^\prime}(\vecx^\prime,t^\prime)}
\nonumber \\
&=& 2 [D(t) \, C^0(t)]_{i j} \updelta^{l l^\prime}\updelta(\vecx -\vecx^\prime) \updelta(t -t^\prime)
\mathcomma \label{eq:noise-zeta}
\end{eqnarray}
where the time dependence of $C^0(t)=C^0(t_0)$ arises from the time dependence of the underlying Hamiltonian and can be determined from the 
{\em equilibrium} fluctuations of the time-independent Hamiltonian $H(t_0)$ with 
$[C^0(t_0)]_{ij}=\int \langle \rho_i(\vecx) \rho_j(0) \rangle_{H(t_0)}\, d^d\vecx$.

In a similar fashion, the source term $\delta r_e$ can be thought of as consisting of deterministic and fluctuating contributions,
\begin{eqnarray}
\delta r_e
&=&
R_e[\delta e,\delta n_a, (\delta e)^2,...]
+ \eta_e
\mathperiod \label{eq:rate-e}
\end{eqnarray}
$R_e$ describes how the local rate of energy creation depends on the deviations $\delta \rho_i = \rho_i -\braket{\rho_i}$
from the (time-dependent) average values,
e.g., high-density regions may respond more strongly to the parameter change than low-density regions.
Again, the linear term is the most relevant one in the expansion
\begin{align}
R_e = -\sum_{j} \alpha_{ej}(t) \delta \rho_j + \landau(\delta\rho^2).
\end{align}
A positive $\alpha_{ee}$ describes a situation, where energy fluctuations are damped. Due to the external drive,  $\alpha_{ee}$ can, however, also be negative as can be checked within the model introduced below in Sec.~\ref{sec:simu}. In this case fluctuations {\em grow} during the parameter change by a factor $\sim e^{|\alpha_{ee}| \tramp}$
which remains, however, finite for $\tramp \to \infty$ as $\alpha_{ee} \sim 1/\tramp$. It is rather straightforward to calculate $\alpha_{ej}$ to leading order in $1/\tramp$ from a microscopic model simply by computing the thermal expectation value $\braket{ \pd{t} H(t) }$ as a function of the average energy and average particle number.
For the noise correlations, we use
\begin{eqnarray}
\braket{\eta_e(\vecx,t) \eta_e(\vecx^\prime,t^\prime)}
&=&
2 A_{ee}(t) \, \updelta(\vecx -\vecx^\prime) \updelta(t -t^\prime)
\mathperiod \label{eq:noise-eta}
\end{eqnarray}
The noise correlations $A_{ee}(t)$ are non-thermal, i.e., they are not determined by a fluctuation-dissipation theorem and should be obtained from a microscopic calculation.
Importantly, 
the time-dependent coefficients $\alpha_{ej}(t)$ and the noise term $\eta_e$ have to scale $\propto \tramp^{-1}$ as they arise from the time-dependent changes of the parameters of the system.
The scaling of $\eta_e$, implies that the noise strength $A_{ee}(t)$ vanishes $\propto \tramp^{-2}$ in the adiabatic limit.

In summary, the general structure of the linear hydrodynamic equations for a quasi-adiabatic parameter change can be written as
\begin{eqnarray}
\pd{t} \drho_i
&=& \sum_j D_{ij}(t) \pdx^2 \drho_j
- \sum_j \alpha_{ij}(t) \drho_j
\nonumber \\
&&
+ \pdx \cdot \veczeta_i
+ \eta_i
\mathcomma \label{eq:linear-eq}
\end{eqnarray}
with the time-dependent diffusion constants $D_{ij}(t)$, the coefficients $\alpha_{ij}(t) \propto \tramp^{-1}$, and the noise correlations as defined in \eqref{eq:noise-zeta} and \eqref{eq:noise-eta}.
For the non-thermal noise we have $A_{ij}(t) =0$ except for the matrix element $A_{ee}(t)$.

Non-linear terms can be neglected for slow enough changes since the deviations from global equilibrium are small. 
This is a non-trivial statement which applies to diffusive systems and can be proven using a straightforward scaling analysis, see \appref{app:scaling}.
Note, however, that linear hydrodynamics cannot be used for systems with momentum conservation in $d=1$ \cite{Spohn1991,Praehofer2004,Mendl2013,Spohn2014}, where the density dependence of the pressure gives rise to non-linearities which are formally relevant and thus cannot be neglected.

\subsection{\label{sec:hyd_lag} Lag of hydrodynamic modes}

The actual fluctuations $\braket{\drho_i(t)\drho_j(t)}\propto C_{ij}(t)$ of hydrodynamic modes lag behind the target fluctuations $C^0_{ij}(t)$.
In the following, we derive the corresponding correlation function in Fourier space.
A Fourier representation with discrete wave vectors $\vecq=(2\pi/L) \,\bm{m}$, $\bm{m} \in \integers^d$, provides a natural parameterization of the non-equilibrium space of hydrodynamic slow modes in linear hydrodynamics as we demonstrate below.
The differences $|C_q(t) -C^0(t)|$ measure the deviation from the ideal adiabatic process giving rise to a finite entropy production of the mode $\vecq$.
The contributions from all modes accumulate to the total entropy production of the system $\Pitot$.

Fourier transformation, $\drho_{i,\vecq} = \int \! \dd^d x \, \e^{-\I \vecq \cdot \vecx}\drho_{i}(\vecx)$,
brings the diffusion equation \eqref{eq:linear-eq} into the standard Langevin form
\begin{eqnarray}
\pd{t} \drho_{i,\vecq}^{\sigma}
&=&
- \sum_j [\gamma_q(t)]_{ij} \drho_{j,\vecq}^{\sigma}
+ \xi_{i,\vecq}^{\sigma}
\mathcomma \label{eq:linear-eq-FT}
\end{eqnarray}
with $\sigma=\pm$ indicating real and imaginary parts,
$\drho_{i,\vecq}^+ = \re\{ \drho_{i,\vecq} \}$, $\drho_{i,\vecq}^- = \im\{ \drho_{i,\vecq} \}$, respectively.
%
The noise terms $\xi_{i,\vecq}^{\sigma}$ collects the contributions from $\veczeta_j$ and $\eta_i$ and has the correlation function
\begin{eqnarray}
\braket{\xi_{i,\vecq}^{\sigma} \xi_{j,\vecq^\prime}^{\sigma^\prime}}
&=&
2 [b_q(t)]_{ij} \, L^d \updelta_{\vecq \vecq^\prime} \updelta^{\sigma \sigma^\prime}
\mathcomma \label{eq:noise-xi}
\end{eqnarray}
with $b_q(t)=\frac{1}{2} [q^2 D(t) C^0(t)+A(t)]$.
The hydrodynamic slow modes $\drho_{i,\vecq}^{\sigma}$ relax
with $q$-dependent rates $\gamma_q(t)= q^2 D(t) + \alpha(t)$.
As a consequence, the time scale of the relaxation vanishes $\propto q^2$ in the long-wavelength limit $q \to 0$ for long enough times.
Note that $\gamma_q(t)|_{|t|\to \infty}= q^2 D(t)|_{|t|\to \infty}$ as $\alpha(t)|_{|t|\to \infty}=0$.
The absence of a minimum time scale leads to power-law behaviors of various observables in hydrodynamic systems.
This also implies the power-law scaling of the entropy production as we will show in \secref{sec:hyd_entropy}.

The linear equation Eq.~\eqref{eq:linear-eq-FT} can directly be solved by
\begin{align}
\drho_{i,\vecq}^{\sigma}(t)
=
\sum_j
\int_{-\infty}^t \! \dd t^\prime \,
[g_q(t,t^\prime)]_{ij} \xi_{j,\vecq}^{\sigma}(t^\prime)\mathcomma  \label{eq:sol}
\end{align}
with the matrix-valued Green function $g_q(t,t^\prime)=\mathcal{T} \e^{-\int_{t^\prime}^{t} \! \dd s \, \gamma_q(s)}$. $\mathcal{T}$ is the time ordering operator which is needed as the matrices $\gamma_q(s)$ do in general not commute when evaluated at different times.
The equal-time fluctuations $C_q(t)$ defined by 
$\braket{\drho_{i,\vecq}^{\sigma}(t)\drho_{j,\vecq^\prime}^{\sigma^\prime}(t)} =\frac{1}{2} [C_q(t)]_{ij} \, L^d \updelta_{\vecq \vecq^\prime} \updelta^{\sigma \sigma^\prime}$ can directly be computed from Eq.~\eqref{eq:sol} and one obtains
\begin{eqnarray}
&&
[C_q(t)]_{ij}
\nonumber \\
&=&
\sum_{mn}
\int_{-\infty}^{t} \! \dd t^\prime \, [g_q(t,t^\prime)]_{im}  [g_q(t,t^\prime)]_{jn} \, 2 [b_q(t^\prime)]_{mn}
\mathperiod \label{eq:actual-corr}
\end{eqnarray}
This equation describes that the system does not adjust instantaneously to $b_q(t)$, thus is always out of equilbrium. This gets more and more severe for small $q$ as relaxation gets slower and slower in this limit.

\subsection{\label{sec:hyd_entropy} Entropy production of\\hydrodynamic modes}
Our prime goal is to find an expression for the total entropy production in the framework of linear fluctuating hydrodynamics.
Our method of calculating the entropy production induced by a quasi-adiabatic parameter change is based on a partition of the system into the hydrodynamic slow modes $\drho_{i,\vecq}^{\sigma}$ and microscopic fast modes.
The entropy balance of the process can be evaluated by adding the contributions in the slow and the fast sector, $\Delta S = \Delta S^{\text{slow}} + \Delta S^{\text{fast}}$.
As we argued earlier in \secref{sec:hyd_eq}, the fast modes are responsible for the noise terms $\veczeta_i$, $\eta_i$.
These modes adjust rapidly to the prescribed protocol as their dynamics occur on the short time scale $\tsc \ll \tramp$.
As discussed in the introductory chapter and in more detail in \appref{app:entr-prod-Boltz}, the relaxation of such fast modes contributes to entropy production with a term proportional to  $\tsc/\tramp$.
As the hydrodynamic modes relax much slower, we have to consider their contribution to entropy production separately.

We consider the Gibbs entropy of the hydrodynamic modes defined as
\begin{eqnarray}
S^{\text{slow}}
& = &
-
\int_{\vecdrho} P(\vecdrho,t) \log\left[ P(\vecdrho,t) \right]
\mathcomma \label{eq:Gibbs-entropy}
\end{eqnarray}
with the normalized probability distribution $\int_{\vecdrho} P(\vecdrho,t)=1$ and $\int_{\vecdrho} \equiv \prod_{i,\vecq,\sigma}\int \dd \drho_{i,\vecq}^{\sigma}$.
We identify two contributions to the change of \eqref{eq:Gibbs-entropy} \cite{Prigogine1967},
\begin{eqnarray}
\dot{S}^{\text{slow}}(t)
& = &
\Phi(t) + \Pi(t)
\mathperiod \label{eq:Gibbs-entropy-decomp}
\end{eqnarray}
Coupling between fast modes and slow modes leads to the entropy flux $\Phi(t)$.
This rate merely refers to an internal redistribution of entropy between different degrees of freedom in the system.
%
%
The entropy production in the slow sector $\Pi(t) >0$ is caused by the irreversible dynamics of the slow modes.
Hence, the total entropy production due to hydrodynamic modes is found as
\begin{eqnarray}
\Pitot
&=&
\int_{-\infty}^{\infty} \! \dd t \,
\Pi(t)
\mathperiod \label{eq:entr-sum-slow-fast}
\end{eqnarray}

As the entropy \eqref{eq:Gibbs-entropy} is defined in terms of the time-dependent probability distribution $P(\vecdrho,t)$,
we turn to the Fokker-Planck equation equivalent to the Langevin equation \eqref{eq:linear-eq},
\begin{eqnarray}
\dfrac{\pp}{\pp t}P(\vecdrho,t)
&=&
-
\sum_{\mu}
\dfrac{\pp}{\pp \drho_{\mu}}\left[a_{\mu}(\vecdrho,t) P(\vecdrho,t)\right]
\nonumber \\
&&
+
\sum_{\mu\nu}
b_{\mu\nu}(t)
\dfrac{\pp^2}{\pp \drho_{\mu} \pp \drho_{\nu}} P(\vecdrho,t)
\mathcomma \label{eq:Fokker-Planck}
\end{eqnarray}
with $a_{\mu} = -\sum_{\nu} \gamma_{\mu\nu} \drho_{\nu}$, $\gamma_{\mu\nu} = [\gamma_q]_{ij} L^d \updelta_{\vecq \vecq^\prime} \updelta^{\sigma \sigma^\prime}$, and
$b_{\mu\nu} = \frac{1}{2}[q^2 D C^0 + A]_{ij} \updelta_{\vecq \vecq^\prime} \updelta^{\sigma \sigma^\prime}$.
For clarity we introduced container indices $\mu = (i,\vecq, \sigma)$.
Using \eqref{eq:Fokker-Planck} in the time derivative of \eqref{eq:Gibbs-entropy} allows us to identify the entropy production rate as
\begin{eqnarray}
\Pi(t)
&=&
\sum_{\mu \nu}
\int_{\vecdrho}
\dfrac{J_\mu(\vecdrho,t) b_{\mu\nu}^{-1}(t) J_{\nu}(\vecdrho,t)}{P(\vecdrho,t)}
\mathcomma \label{eq:ent-prod-start}
\end{eqnarray}
with the probability current
\begin{eqnarray}
J_\mu(\vecdrho,t)
&=&
a_{\mu}(\vecdrho,t) P(\vecdrho,t)
\nonumber \\
&& 
-\sum_{\nu} b_{\mu\nu}(t) \dfrac{\pp P(\vecdrho,t)}{\pp \drho_{\nu}}
\mathperiod \label{eq:prob-current}
\end{eqnarray}
The straightforward derivation of \eqref{eq:ent-prod-start} was discussed by Tom\'{e} \cite{Tome2006} (for the case $b_{\mu\nu}\propto \updelta_{\mu\nu}$).
The intermediate steps are given in \appref{app:multi-FP-rates} for completeness.
In the further evaluation of \eqref{eq:ent-prod-start} we make use of the fact that the Fokker-Planck equation is solved by a Gaussian probability distribution
$P(\vecdrho,t) \propto \e^{-\frac{1}{2} \sum_{\mu\nu} \drho_{\mu} [C^{-1}(t)]_{\mu \nu} \drho_{\nu}}$
with the time-dependent correlation matrix $C_{\mu \nu}(t) \equiv \frac{1}{2}[C_q(t)]_{ij} L^d \updelta_{\vecq \vecq^\prime} \updelta^{\sigma \sigma^\prime}$ given in \eqref{eq:actual-corr}.
The Gaussian integral yields:
\begin{eqnarray}
\Pi(t)
&=&
\dfrac{L^d}{(2\pi)^d}\int \! \dd^d q \,
f[\gamma_{q,t}, b_{q,t}, C_{q,t}]
\mathperiod \label{eq:ent-prod-trace}
\end{eqnarray}
with $f[\gamma_{q,t}, b_{q,t}, C_{q,t}]= \tr\{ \gamma_{q,t}\trans b_{q,t}^{-1}\gamma_{q,t} C_{q,t} + C_{q,t}^{-1}b_{q,t}  -2\gamma_{q,t}\}$.
We performed the continuum limit in the final expression to obtain the integral in Fourier space.
%
%
%
The total entropy due to hydrodynamic modes produced by the parameter change is given by
\begin{eqnarray}
\Pitot
&=&
\dfrac{L^d}{(2\pi)^d}
\int_{-\infty}^{\infty} \! \dd t
\int \! \dd^d q \,
f\left[\gamma_{q,t}, b_{q,t}, C_{q,t}\right]
\mathcomma \label{eq:ent-prod-tot}
\end{eqnarray}
respectively.

An evaluation of Eq.~\eqref{eq:ent-prod-tot} involves the computation of the correlation function $C_{q,t}$ from Eq.~\eqref{eq:actual-corr} which needs the knowledge of the Green function $g_q(t,t')$  for a set of coupled diffusion equations with time-dependent parameters. Quantitatively, such a problem can only be solved numerically.
To obtain qualitative analytical results in the limit of large $\tramp$, it is useful to perform a scaling analysis
by introducing the dimensionless variables $\tilde{t}=t/\tramp$ and $\tilde{q}=(\diffunit \tramp)^{1/2} q$.
Here, the constant $\diffunit$ carries the unit of $D_{ii}$.
We proceed by performing first a naive scaling analysis of all terms contributing to the entropy production, Eq.~\eqref{eq:ent-prod-tot}.
Introducing the rescaled damping coefficient $\tilde{\gamma}(\tilde{q},\tilde{t})=\tramp \,\gamma[(\diffunit\tramp)^{-1/2} \tilde{q},\diffunit\tramp \tilde{t} ]$ and the rescaled correlation matrices
$\tilde{b}(\tilde{q},\tilde{t})=\tramp \, b[ (\diffunit\tramp)^{-1/2} \tilde{q},\diffunit\tramp \tilde{t}]$,
$\tilde{C}(\tilde{q},\tilde{t})= C[(\diffunit\tramp)^{-1/2} \tilde{q},\diffunit \tramp \tilde{t} ]$
leads to
\begin{subequations} \label{subeq:ent-prod-tot-scaled}
\begin{eqnarray}
\Pitot
&=&
F \,
\dfrac{L^d}{(\diffunit \tramp)^{d/2}}
\mathcomma \label{eq:ent-prod-tot-scaled}
\\
F
&=&
\int_{-\infty}^{\infty} \! \dd \tilde{t}
\int \! \frac{\dd^d \tilde{q}}{(2 \pi)^d} \,
f\left[\gamma_{\tilde{q},\tilde{t}}, b_{\tilde{q},\tilde{t}}, C_{\tilde{q},\tilde{t}}\right]
\mathperiod \label{eq:ent-prod-tot-scaled-prefactor}
\end{eqnarray}
\end{subequations}
Hence, naive scaling predicts that the adiabatic limit is reached algebraically with $\Pitot \propto \tramp^{-d/2}$ in $d$ spatial dimensions.
The dimensionless prefactor $F$ depends on the details of the protocol encoded in the functions $D_{ij}(\tilde{t})/\diffunit$, $\tilde{C}^0_{ij}(\tilde{t})$, and $\tilde{\alpha}_{ij}(\tilde{t})$.
To obtain the scaling of $b(q,t)$ to leading order
we used that
the non-thermal noise $\eta_i$ only contributes
subleading corrections, see \appref{app:scaling}.

However, the naive scaling relies on the assumption that $F$ converges to a $\tramp$ independent value for $\tramp \to \infty$.
This is only the case if the integrals defining $F$ are convergent and thus independent of a cutoff to the $q$ modes.
The time integral is harmless as for $t \gg \tramp$ or, equivalently,  $\tilde t \gg 1$ the entropy production vanishes.
As we discuss in \appref{app:entr-prod-hyd-dim}, the $\tilde q$ integrand is proportional to $\tilde q^{-2}$ for large $\tilde q$ .
Therefore, the integral converges only in $d=1$.
In $d\ge 2$, large momenta of the order of the cutoff scale contribute. 
In rescaled variables this cutoff is given by 
$(\diffunit \tramp)^{1/2} \Lambda$, where the momentum cutoff $\Lambda \sim 1/\ell$ is of the order of the mean-free path $\ell$ of the system. 
As a consequence, the power-law exponent in \eqref{eq:ent-prod-tot-scaled} is altered.
As we show in \appref{app:entr-prod-hyd-dim}, the leading-order contributions to the entropy production arising from hydrodynamic modes are
\begin{eqnarray}
\frac{\Pitot}{L^d}
&\sim&
\begin{cases}
(\mathcal D \tramp)^{-1/2} & d=1  \\
(\mathcal D \tramp)^{-1}\log\left( \Lambda^2 \diffunit \tramp\right) & d=2   \\
(\mathcal D \tramp)^{-1} \Lambda^{d-2} & d > 2 
\end{cases}
\mathperiod \label{eq:ent-prod-tot-scaled-cutoff}
\end{eqnarray}
In $d=3$, where high-$q$ modes dominate, we recover the result $\Delta S\propto \tramp^{-1}$ of the Boltzmann equation, see \appref{app:entr-prod-Boltz}. This is not surprising 
as high-$q$ modes relax exponentially with a relaxation time of order $\tsc$ for $q\sim 1/\ell$.
One should, however, note that in this case the hydrodynamic description predicts the wrong prefactor as the hydrodynamic description may cover only a small fraction of all modes of the system.
In contrast, in $d\le 2$ entropy production is dominated by hydrodynamics. While in $d=2$ there is only a logarithmic enhancement of little importance, hydrodynamic long-time tails lead to an extremely slow decay of entropy production, $\Delta S \sim \tramp^{-1/2}$ in $d=1$.




\begin{figure*}[t]
\centering
\begin{minipage}{246pt}
\begin{flushright}
\includegraphics{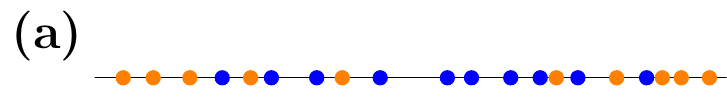}
\includegraphics{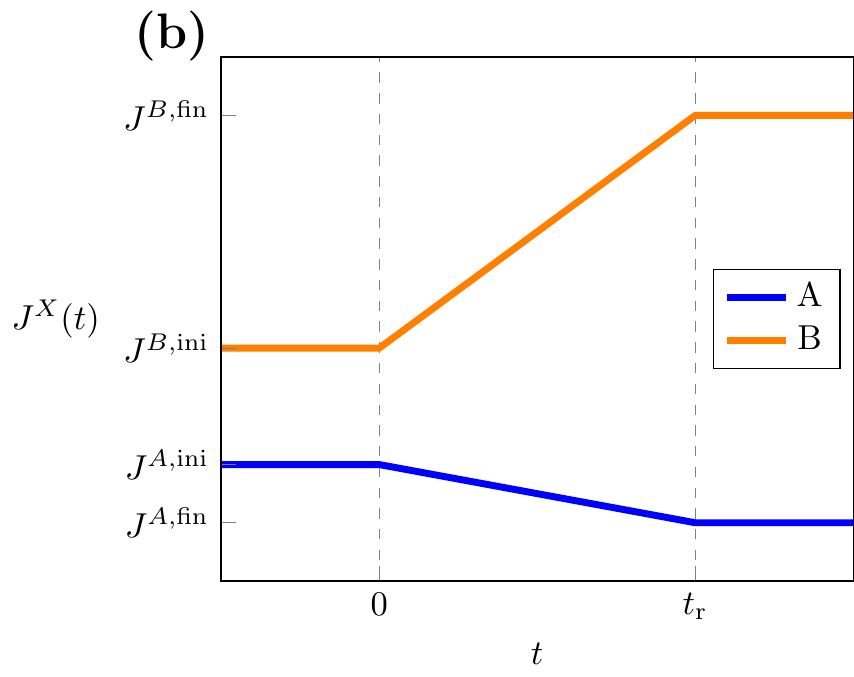}
\end{flushright}
\end{minipage}
\hfill
\begin{minipage}{246pt}
\includegraphics{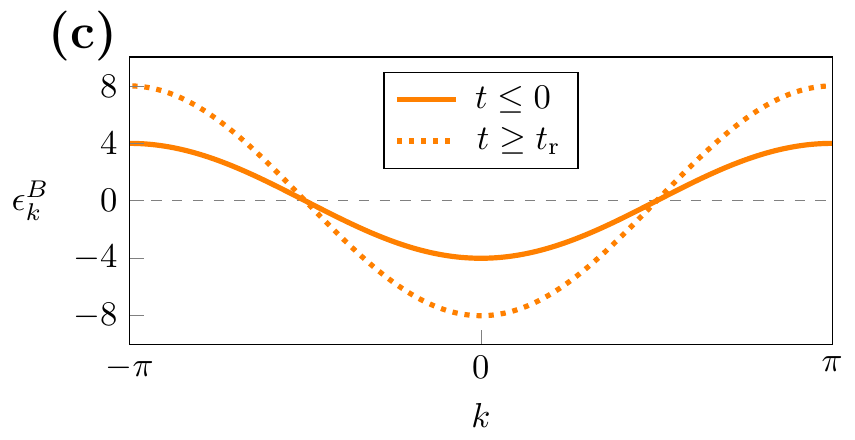}
\includegraphics{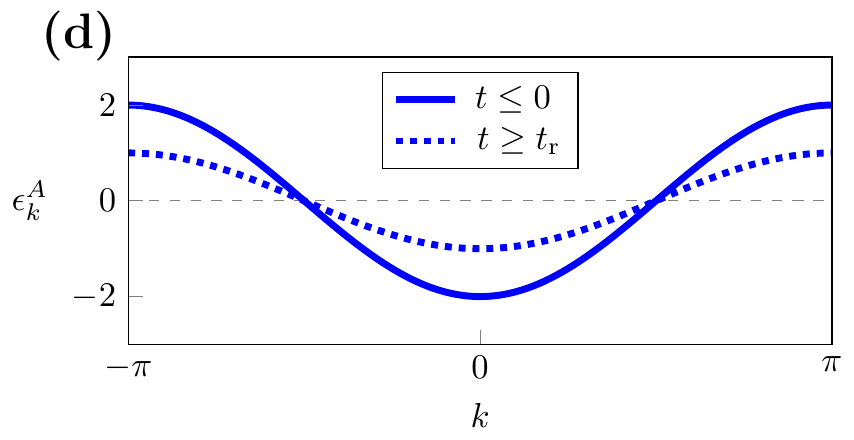}
\end{minipage}
\caption{\label{fig:tuning-Js} 
(a) Illustration of a typical configuration of particle species A and B in real space.
(b) The hopping parameters of particle species A and B are changed linearly from $J^{X,\text{ini}}$ to $J^{X,\text{fin}}$ ($X=A,B$) in time $\tramp$.
(c,d) The band width of particle type A (B) is decreased (increased) in the parameter change.}
\end{figure*}

\subsection{\label{overview1D} Overview over entropy production in $d=1$}
We can now connect the results obtained from the hydrodynamic calculations with the result from the Boltzmann equation.
For simplicity we focus on $d=1$ where the effects are most pronounced. The following discussion builds on the assumption that a single scattering time $\tsc$ dominates all physics properties. More precisely, we assume that $\tsc$ changes only by factors of order $1$  during the parameter change. Furthermore, we assume that the transport scattering times for energy and particle number transport are similar and that these transport scattering times  differ only by factors of order $1$ from, e.g., the time-scale which describes how inelastic scattering processes  relax  distribution functions towards equilibrium. These conditions are met by the model considered in Sec.~\ref{sec:simu} below but can also be violated, e.g., due to proximity to some integrable model \cite{Jung2006} or when disorder leads to localization.

The Boltzmann equation predicts $\frac{\Pitot}{N} \sim \frac{\tsc}{\tramp}$ for the entropy production per particle, see \appref{app:entr-prod-Boltz},  while we find in the hydrodynamic regime $\frac{\Pitot}{N} \sim \frac{a}{\sqrt{\mathcal D \tramp}}=\frac{\tau_0}{\sqrt{\tsc \tramp}}$, where $a$ is the distance of particles and we introduce the timescale $\tau_0=a/v$ obtained from the distance $a$ and the average velocity $v$ of the particles with $\mathcal D=v^2 \tsc$.  Using this result, we obtain for $\tsc \gg \tau_0$ three different regimes 
\begin{eqnarray}
\Pitot
&\sim&
S_{\rm quench}  \begin{cases}
1 \qquad &  \tsc \gtrsim \tramp \\
\frac{\tau}{\tramp} & \tsc \lesssim \tramp \lesssim \frac{\tsc^3}{\tau_0^2} \\
\frac{\tau_0}{\sqrt{\tsc \tramp}} & \frac{\tsc^3}{\tau_0^2} \lesssim \tramp
\end{cases}
\mathperiod \label{eq:ent-prod-overview}
\end{eqnarray}
When the parameters are changed rapidly on the time scale set by the scattering,  $\tramp \lesssim \tsc$, the distribution function of quasiparticles does not adjust during the time when the  Hamiltonian is changed and we expect that the entropy production is approximately given by the result for a quench, $\tramp=0$. For $\tsc \lesssim \tramp \lesssim \frac{\tsc^3}{\tau_0^2}$ the entropy production by high-energy modes dominates and the entropy production scales with $1/\tramp$. For $\tramp \gtrsim  \frac{\tsc^3}{\tau_0^2}$ the hydrodynamic long-time tails dominate with $\Pitot \sim 1/\sqrt{\tramp}$.

\section{\label{sec:simu} Numerical evidence of hydrodynamic scaling}

%

\subsection{\label{sec:simu-model} Two-component-gas model}


To verify the power-law scaling $\propto t^{-1/2}$ of the entropy production in one dimension,
we consider a simple toy model \cite{Sachdev1997,Damle1998,Garst2001,Lux2014}:
a 1D classical gas of two different particle species A and B as illustrated in \figref{fig:tuning-Js} (a). 
The kinetic energy of the two-component gas is
\begin{eqnarray}
H_{\text{kin}}
&=&
\sum_{i=1}^{N_A} \dis{A}{k_i}(t) + \sum_{j=1}^{N_B} \dis{B}{k_j}(t)
\mathcomma \label{eq:Hamiltonian-kinetic}
\end{eqnarray}
where $k_i$ denotes the momentum of particle $i$ and $N_{A,B}$ are the particle numbers.
Interactions are taken into account by elastic hard-core collisions. While a collision of two classical particles with the
same dispersion does not change the momentum-distribution function (as their momenta are simply exchanged), the two-body scattering
of particles with different dispersion leads to equilibration \cite{Lux2014}. 

In this setup the mean free path $\ell$ is approximately given by the average distance $a$ of A and B particles, $\ell \approx a$. Similarly,
the scattering time $\tau=\ell/v$ is of the order of $\tau_0=a/v$ where $v$ is the typical velocity of particles, $\tau \approx \tau_0$. 

To be able to tune the ratio of the scattering length $\ell$ and the distance of particles $a$ or, equivalently, of $\tau$ and $\tau_0$, we furthermore introduce a `tunneling probability' $\Gamma$, $0\le \Gamma \le 1$: when two particles meet, they pass each other with the probability $\Gamma$,  while they scatter with the probability $1-\Gamma$. Therefore, the scattering time for finite $\Gamma$ is given by
\begin{align}
\taugamma \approx \frac{\tau_{0}}{1-\Gamma}\mathperiod \label{tauGamma}
\end{align}
 For our simulations, we determine for each $\Gamma$ the scattering time $\taugamma=\delta t \frac{N_A}{N_c}$ in the initial thermal state ($t<0$) by counting the number of collisions per A-particle, $N_c/N_A$, in a time-interval
 $\delta t$.

The dispersion of the particles
\begin{eqnarray}
\dis{X}{k}(t)
&=&
- 2 J^{X}(t) \cos(k) 
\mathcomma \label{eq:dis_AB}
\end{eqnarray}
for particles of type $X=A,B$, respectively, is time-dependent and changes slowly.
The $\cos$ dispersions can be understood as being derived form a lattice Hamiltonian with lattice constant $1$ and time-dependent hopping parameters $J^{X}(t)$.
The particle momenta are restricted to the Brillouin zone $-\pi < k \leq \pi$.
Thus, umklapp scattering between particles of type A and B occurs which violates momentum conservation and leads to a diffusive behavior \cite{Lux2014}.

We consider a linear ramp of the hopping parameters in time $\tramp$
as defined by the functions
\begin{subequations} \label{subeq:hopping}
\begin{eqnarray}
J^{X}(t)
&=&
J^{X,\text{ini}} \left[ 1 -h(t)\right]
+
J^{X,\text{fin}} h(t)
\mathcomma \label{eq:hopping}
\\
h(t)
&=&
\begin{cases}
0 \mathcomma & t < 0 \\
t/\tramp \mathcomma & 0 \leq t \leq \tramp \\
1 \mathcomma & t > \tramp
\end{cases}
\mathperiod \label{eq:hopping-func}
\end{eqnarray}
\end{subequations}
The ramp protocol is illustrated in \figref{fig:tuning-Js} (b)--(d). 
The numerical results presented in the subsequent section \secref{sec:simu-micro} are obtained for $J^{A,\text{fin}} = 0.5 \, J^{A,\text{ini}}$, $J^{B,\text{fin}} = 2\, J^{B,\text{ini}}$ and
$J^{A,\text{ini}}=1$, $J^{B,\text{ini}}=2$.
%
Importantly,  these parameters avoid a special integrable point obtained when the two hopping rates are equal, $J^{A}(t^\ast)=J^{B}(t^\ast)$ \cite{Garst2001}.
The ramp protocol is continuous but non-analytic at $t=0$ and $t=\tramp$. We have checked that such a non-analytic dependence does not invalidate our analytic predictions for the entropy production to leading order in $1/\tramp$.

From the viewpoint of hydrodynamics, our model possesses for $\Gamma >0$ three diffusive modes for energy conservation and the number conservation of each particle species and is thus described by Eqs.~\eqref{eq:cont-e}--\eqref{eq:current-n}. As discussed above, energy conservation is weakly violated due to the slow change of the dispersion.
If we consider only hard-core collisions by setting $\Gamma=0$, the situation is a bit more complicated. As $A$ particles cannot pass $B$ particles, diffusion of the difference of densities, $\rho_A-\rho_B$, is prohibited. Furthermore,  our model also has extra `hidden' conservation laws:  a given sequence of particles $ABAAB\dots$ does not change as function of time. Such a non-local `string order' is not expected to give rise to further hydrodynamic modes. 
As we show below, we find numerically for the model with $\Gamma=0$ the same asymptotic behavior for entropy production 
as in the models with $\Gamma>0$.


%
%
%

%




\begin{figure*}[t]
\centering
\hfill
\begin{minipage}{245.9pt}
\includegraphics{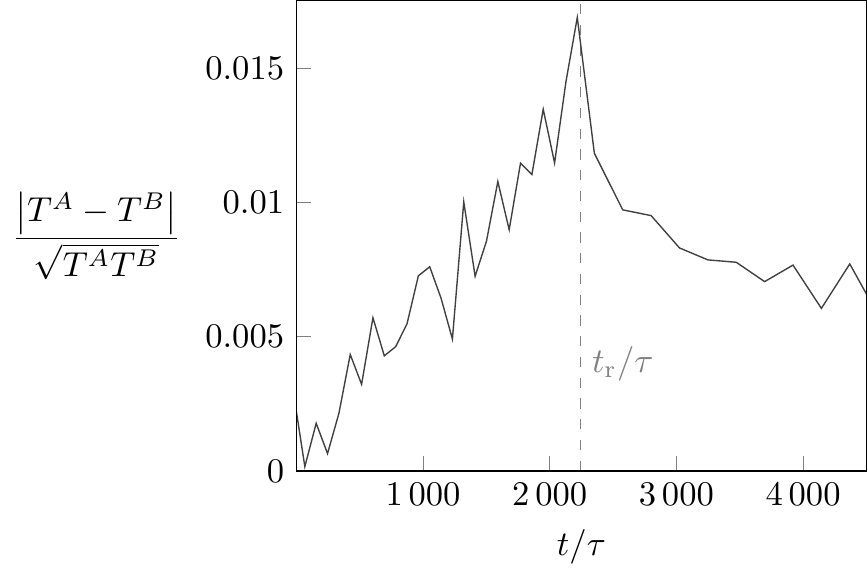} 
\end{minipage}
\hspace{-25pt}
\begin{minipage}{245.9pt}
\includegraphics{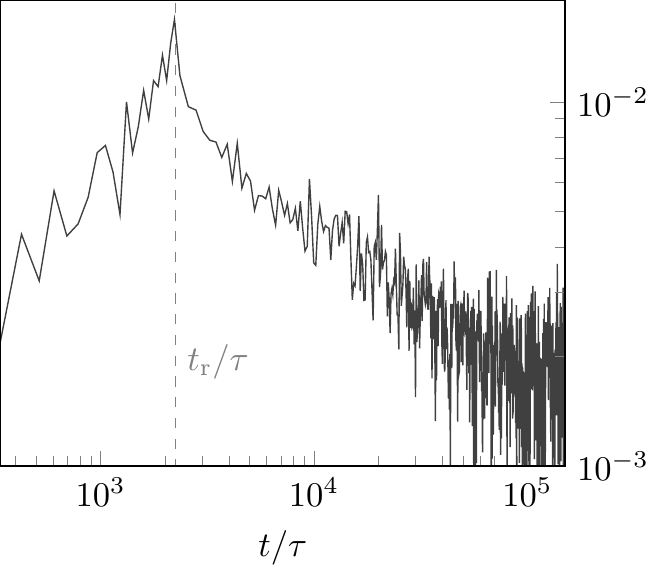} 
\end{minipage}
\hfill
\caption{\label{fig:relax-example} 
The normalized temperature difference $\frac{|T^A -T^B|}{\sqrt{T^A T^B}}$ of the two subsystems (obtained by solving Eqs.~\eqref{eq:inversion-N} and \eqref{eq:inversion-E}, see text) 
can be used to track how far the system is out of equilibrium (left figure: linear scale, right figure: logarithmic scale). In the shown example, parameters change very slowly according to the quench protocol defined in \figref{fig:tuning-Js} with $\tramp\approx 2200\, \taugamma$, where $\taugamma$ is the average time between two AB scattering events of a single particle ($N_A=N_B=10^5/2$, $\Gamma=0$). During the parameter change, $0\le t \le \tramp$, the temperature difference grows linearly in time and relaxes slowly for $t \gg \tramp$. For $t\to \infty$, an equilibrium state with $T^A=T^B$ is reached.
} 
\end{figure*}

\begin{figure}[t]
\includegraphics{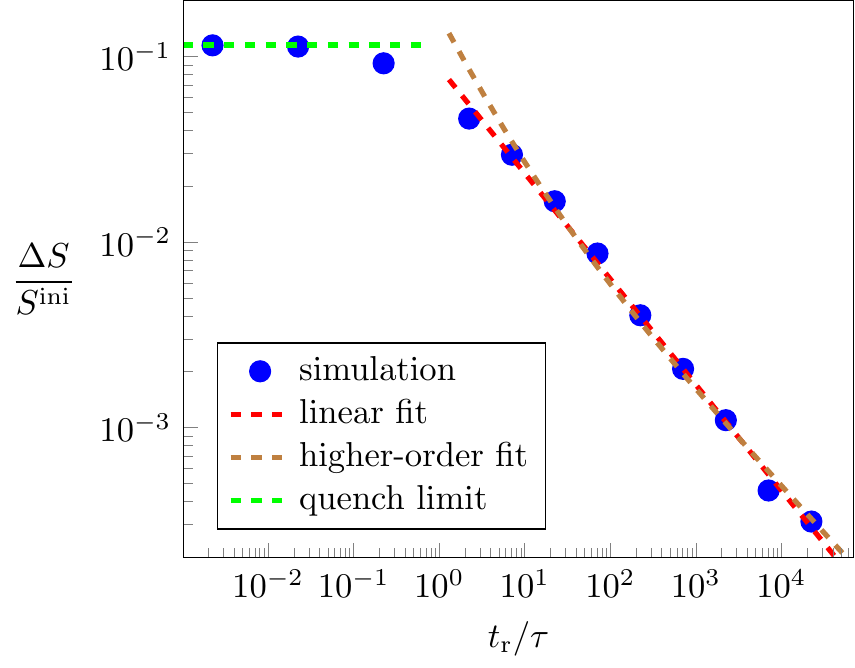}
\caption{\label{fig:main-result} 
Our numerical simulation ($\Gamma=0$ and all parameters as in \figref{fig:relax-example}) shows
that the entropy production of the two-component gas decays algebraically for large quench times $\tramp \gg \taugamma$ (blue dots).
The red-dashed line shows a linear fit, $\Delta S \sim t^{-0.57}$, slightly deviating from the prediction of linear hydrodynamics, $\Delta S \sim \tramp^{-1/2}$. The brown-dashed line shows the fit, $\frac{\Delta S}{S\ini} \approx c_0 \,(\taugamma/\tramp)^{-1/2} + c_1\,(\taugamma/\tramp)^{-1}$ ($c_0 = 0.047$, $c_1 = 0.120$), which includes subleading corrections. The green dashed line denotes the entropy production for a sudden change of parameters which can be obtained analytically, see text.
}
\end{figure}

\subsection{\label{sec:simu-micro} Simulation}

We prepare the particles in a homogeneous initial state at the inverse temperature $\beta^{\text{ini}}= 1$.
The momenta are distributed according to the thermal distribution $f^{0,X} = \e^{-\beta^{\text{ini}} \dis{X,\text{ini}}{k}}/z^{X}$ with the fugacities $z^{X}$.
The positions in real space are drawn from a uniform distribution.
Our simulation is based on the bookkeeping algorithm used in \cite{Lux2014} to study the emergence of hydrodynamic long-time tails after a sudden interaction quench in the 1D Bose-Hubbard model.
Here, we introduce the time-dependence of the dispersion \eqref{eq:dis_AB} as a new feature.
The linear ramp introduced in Eq.~\eqref{subeq:hopping} has the advantage that it allows to calculate analytically, when a future collision
of neighboring particles occurs. The algorithm consist of keeping a list of such future collisions, which is updated after each scattering event.
Here, the momenta of the scattered particles are determined by the conservation of energy and lattice momentum.
The algorithm can easily be solved for the dynamics of $10^5$ particles, tracking about $10^{10}$ collisions.

The entropy of the classical particles can be obtained from 
\begin{subequations} \label{subeq:entropy-micro}
\begin{eqnarray}
S
&=& S^A + S^B
\mathcomma \label{eq:entropy-micro}
\\
S^{X}
&=&
-
\dfrac{L}{2 \pi}
\int_{-\pi}^{\pi}
\! \dd k \,
f^{X}_k \log\left(f^{X}_k \right)
\mathperiod \label{eq:entropy-micro-AB}
\end{eqnarray}
\end{subequations}
$f^{X}_k$ denotes the momentum distributions of particle types $X=A,B$, respectively.
This formula has, however, two disadvantages for practical calculations. First, it requires a numerical calculation of the distribution function
$f^X_k$ by some binning procedure which introduces some (small) numerical error. More importantly, the system relaxes very slowly $\sim 1/\sqrt{t}$ to its final thermal state for $t \gg \tramp$. Thus stopping the simulation at some finite time $t\gg \tramp$ introduces a substantial error.
For a high-precision determination of the final entropy, we therefore use a trick which is based on two observations: (i) For $t > \tramp$ the energy is exactly conserved, $E(t>\tramp)=E(\tramp)=\text{const.}$, and (ii) for $t \to \infty$ the system reaches a thermal state. The temperature of this final state can easily be computed from the knowledge of $E(\tramp)$. Here one uses that the interaction energy of the point particles is exactly zero, therefore the thermodynamics of a gas of classical point-particles scattering elastically is exactly given by the thermodynamics of a non-interacting gas. One can easily compute the entropy difference of the initial state and the final state for $t\to \infty$, using a thermal distribution function, $f^{0}_k(\beta,z^{A/B},J^{A/B}) = \e^{-\beta \dis{A/B}{k}}/z^{A/B}$ with the inverse temperature $\beta$ and the fugacities $z^{A/B}$  obtained from
\begin{subequations}
\begin{eqnarray}
E&=& \sum_{X=A,B} E^X \mathcomma \\
E^X &=&
\dfrac{L}{2 \pi}
\int_{-\pi}^{\pi}
\! \dd k \,
\dis{X}{k}
f^{0}_k(\dis{X}{},\beta^X,z^X)
\mathcomma \label{eq:inversion-N}
\\
N^X
&=&
\dfrac{L}{2 \pi}
\int_{-\pi}^{\pi}
\! \dd k \,
f^{0}_k(\dis{X}{},\beta^X,z^X)
\mathcomma\label{eq:inversion-E}
\end{eqnarray}
\end{subequations}
with $X=A,B$ and $\beta^A=\beta^B=\beta$. The entropy of final and initial state is then calculated from
\begin{eqnarray}
S
&=& \sum_{X=A,B}
\beta\, E^X
-
N^X \log(z^X)
\mathcomma \label{eq:entropy-micro-eq}
\end{eqnarray}
which is equivalent to Eq.~\eqref{subeq:entropy-micro} for equilibrium systems.

One way to track how far the system is out of equilibrium is to define two separate temperatures $T^A(t)$ and $T^B(t)$ at each time for the two subsystems $A$ and $B$
by tracking the energies $E^A$ and $E^B$ and by solving Eqs.~\eqref{eq:inversion-N} and \eqref{eq:inversion-E} for $\beta^X=1/T^X$. 
$T^A$ and $T^B$ are not true thermodynamic temperatures, but the normalized difference $\frac{|T^A-T^B|}{\sqrt{T^A T^B}}$ is a useful measure
to track how far the system is out of equilibrium.
As shown in \figref{fig:relax-example}, the temperature difference grows approximately linearly in $t$ during the parameter change, $0\le t\le \tramp$.
Taking into account that for the example shown in \figref{fig:relax-example}
each particle undergoes collisions of the order of $10^3$ during the parameter change, the dimensionless temperature difference of $10^{-2}$
is relatively high.
For $t >\tramp$, the temperature difference relaxes very slowly back to zero. Fluctuating hydrodynamics predicts 
$|T^A-T^B|\sim 1/\sqrt{t}$ in this regime \cite{Lux2014}, consistent with our numerics.
Note, however, that the 
entropy production is {\em not} proportional to $(T^A-T^B)^2\sim 1/t$ as one would expect from the coupling of two thermal reservoirs as such
a large entropy production rate would lead to a divergence of the total change of entropy when integrated over time. This shows that the
local temperatures do not correctly represent the non-equilibrium state.

In \figref{fig:main-result} we show the central result of our numerical study: the entropy production
 $\Pitot(\tramp)$ as a function of the ramp time $\tramp$ measured in units of the single-particle collision time.
The entropy production for a sudden change of parameters, i.e., a quench with $\tramp \ll \taugamma$, can be calculated analytically
using that the distribution function does not change for $t\ll \taugamma$, which allows to calculate the finial energy using the initial distribution function. For our protocol, we find that the entropy increases by $12\,\%$ in this limit. As expected, the entropy production is reduced when
parameters change more slowly. For $\tramp \gtrsim 10 \taugamma $, the entropy production can approximately be described by a power law, $\Delta S \sim \tramp^{-\plexp}$, as can be seen on the
double-logarithmic plot of  \figref{fig:main-result}.
A linear fit yields $\plexp\approx 0.57$.
We attribute the apparent deviation from the predicted value $\plexp=1/2$ to subleading corrections beyond linear hydrodynamics, discussed in \appref{app:scaling}.
A fit with the function $c_0 \,(\taugamma/\tramp)^{-1/2} + c_1\,(\taugamma/\tramp)^{-1}$ describes the data also very well (see \figref{fig:main-result}), and yields the prefactors $c_0 = 0.047$ and $c_1 = 0.120$.
Thus, the nominally subleading correction can distort the leading-order behavior on intermediate time scales.
Nevertheless, we conclude that the simulation results are clearly consistent with our analytical approach.

\begin{figure*}[t]
\centering
\hfill
\begin{minipage}{245.9pt}
\includegraphics{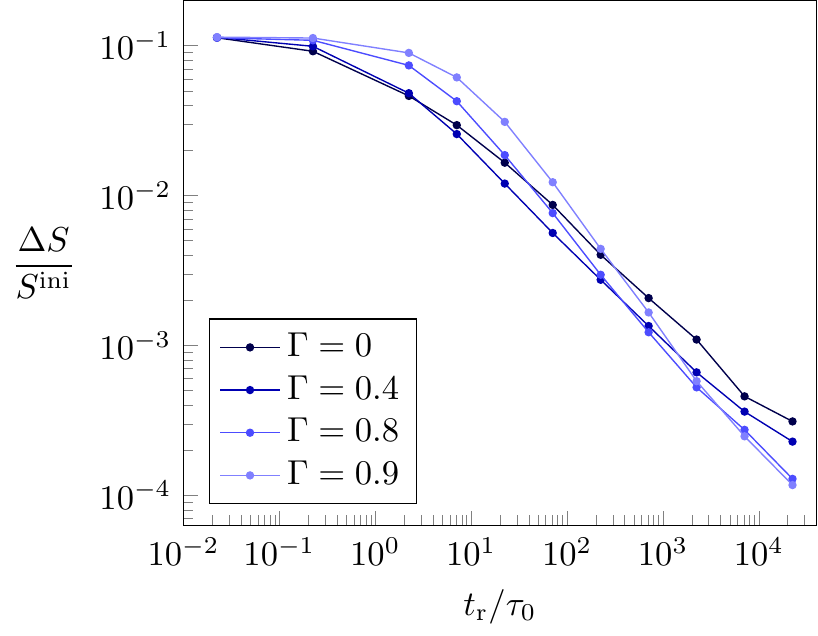} 
\end{minipage}
\hspace{-10pt}
\begin{minipage}{245.9pt}
\includegraphics{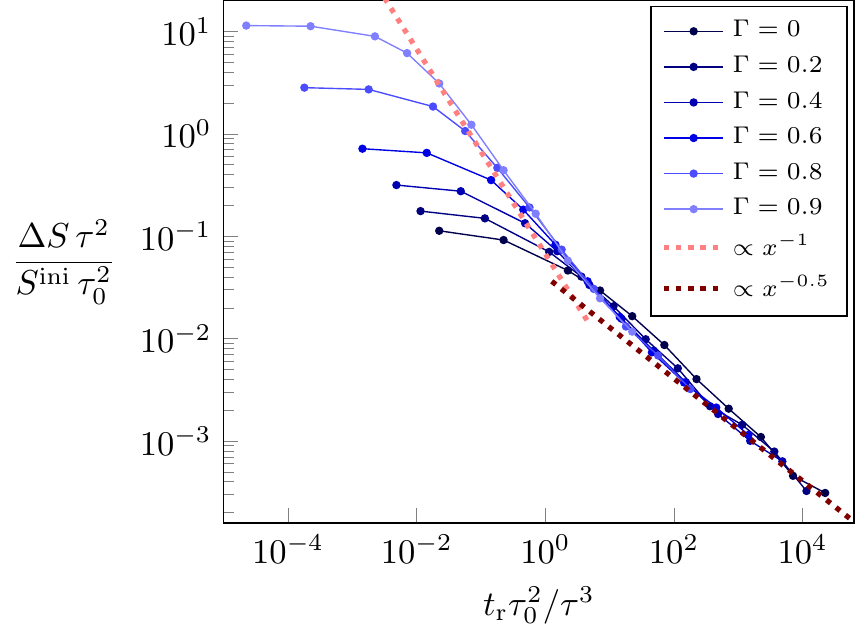} 
\end{minipage}
\hfill
\caption{\label{fig:result-tunnel} 
Entropy production in a model with a finite tunneling rate $\Gamma$ (all other parameters chosen as in \figref{fig:main-result}, $\tau_0=\tau(\Gamma=0)$ is obtained from the scattering time for $\Gamma=0$). Left panel: At intermediate times, entropy production is enhanced when the scattering time $\taugamma\approx \tau_0/(1-\Gamma)$ gets larger. In contrast, for long times, the entropy production is reduced. Right panel: A scaling plot shows that the analytical estimates of Eq.~\eqref{eq:ent-prod-overview} are valid, see text. The dashed lines show that the rescaled entropy production $\Delta S \,\tau^2/(S\ini \tau_0^2)$ scales with $1/x$ at intermediate times and with $1/\sqrt{x}$ in the long-time limit for $x=\tramp \tau_0^2/\tau^3$.
} 
\end{figure*}

To investigate whether our predictions for prefactors and for the interplay of hydrodynamic and other modes discussed in Eq.~\eqref{eq:ent-prod-overview} are valid, we study the entropy production for our model in the presence of a finite tunneling rate in \figref{fig:result-tunnel}.
Most importantly, we can use the parameter $\Gamma$ to increase the scattering time $\taugamma$ and to vary the ratio of $\tau_0$ and $\taugamma$ according to Eq.~\eqref{tauGamma}.

The Boltzmann equation predicts that the entropy production increases for larger  $\taugamma$ while our hydrodynamic theory shows the opposite effect.
This is reflected in the left panel of   \figref{fig:result-tunnel}.
For larger $\Gamma$ and thus larger $\taugamma$ the entropy production increases for short times but is reduced in the long-time limit where hydrodynamics dominates. To extract the different regimes described by Eq.~\eqref{eq:ent-prod-overview} we use a scaling plot in the right panel of \figref{fig:result-tunnel}.
We consider a rescaled entropy production $y=\frac{\Delta S}{S\ini} \frac{\taugamma^2}{\tau_0^2}$ which is plotted as function of the scaling variable $x=\frac{\tramp \tau_0^2}{\taugamma^3}$. According to Eq.~\eqref{eq:ent-prod-overview} we expect (for $\tramp \gg \taugamma$) a crossover from the Boltzmann regime $y \sim 1/x$ for $x\ll 1$ to the hydrodynamic regime $y \sim 1/\sqrt{x}$
for $x\gg 1$. This is fully confirmed by our numerical data.

\section{\label{sec:conc} Conclusions}
When the initial state is prepared in a typical ultracold-atom experiment, changing parameters slowly is important to keep entropy production at
a minimum while approaching the desired state of matter. In a harmonic trap, the initial and final state will typically have a very different distribution of atoms and energy. Reaching the final thermal state is therefore associated with the transport of energy and atoms over a length scale set by the radius of the cloud. These transport processes generate entropy and are a major bottleneck for isentropic state evolution \cite{Rapp2010}. 

As our study shows, a similar bottleneck for isentropic processes arises even in translationally invariant systems, where there is no need to transport energy or particles over large distances. Here the bottleneck is the buildup of fluctuations of conserved quantities which characterize each thermal state. The longer the wavelength of these fluctuations, the longer it takes to equilibrate them.  Our study has shown
that in diffusive systems this is the limiting factor for entropy production in dimensions $d\le 2$. The effect is most pronounced in $d=1$ where entropy production scales with $1/\sqrt{\tramp}$ when parameters change slowly on the time scale $\tramp$. In $d=2$ there is only a logarithmic enhancement of the $1/\tramp$ dependence of entropy production expected for $d>2$. 

To reduce entropy production in a slow process, the Boltzmann equation predicts that it is best to have a short scattering time so that the system remains always close to equilibrium. Remarkably, the situation is opposite in the hydrodynamic regime, especially in $d=1$. Here, a larger diffusion constant -- obtained for large scattering times -- helps to build up thermal fluctuations. Therefore, it is better to have large scattering times in this regime to reduce entropy production. We can also determine the optimal scattering time $\taugamma$ to achieve the minimal entropy production for a fixed time scale $\tramp$ on which the Hamiltonian is modified. According to Eq.~\eqref{eq:ent-prod-overview} the optimal relaxation time $\tau$ is given by $(\tramp \tau_0^2)^{1/3}$ which leads to an entropy production proportional to $(\tau_0/\tramp)^{2/3}$.

%

An interesting open question is the entropy production arising from hydrodynamic modes in models with momentum conservation, discussed briefly in \appref{app:mom-con}. In this case non-linear terms in the hydrodynamic equations turn out to be {\em relevant} (in the renormalization group sense). They change the exponents in thermal correlation functions \cite{Mendl2013,Spohn2014} in dimensions $d < 2$ (with logarithmic corrections in $d=2$) and are thus expected to also modify the exponents describing entropy production, especially in $d=1$.

\begin{acknowledgments}
The numerical simulations were performed on the HPC system CHEOPS at RRZK, University of Cologne, funded by DFG through Grant No. INST 216/512/1FUGG.
Furthermore, this work was supported by CRC 1238 (project C04, project number 277146847).
\end{acknowledgments}

\appendix

\section{\label{app:multi-FP-rates} Separating rates of entropy production and entropy flux using the multivariate Fokker-Planck equation}

Our derivation of the entropy production rate of hydrodynamic modes $\Pi(t)$ closely follows the discussion in Ref.~\cite{Tome2006}.
We take the time derivative of the Gibbs entropy \eqref{eq:Gibbs-entropy},
\begin{eqnarray}
\dot{S}^{\text{slow}}
& = &
-
\int_{\vecdrho} \pd{t} P(\vecdrho,t) \log\left[ P(\vecdrho,t) \right]
\mathcomma \label{eq:Gibbs-entropy-derivative}
\end{eqnarray}
and insert the Fokker-Planck equation \eqref{eq:Fokker-Planck} in the form of a continuity equation (reflecting the conservation of probability),
\begin{eqnarray}
\dfrac{\pp}{\pp t}P(\vecdrho,t)
&=&
-
\sum_{\mu}
\dfrac{\pp}{\pp \drho_{\mu}} J_\mu(\vecdrho,t)
\mathperiod \label{eq:Fokker-Planck-continuity}
\end{eqnarray}
The probability current $J_\mu(\vecdrho,t)$ is defined in \eqref{eq:prob-current}.
We integrate by parts and use that $ 
\frac{\pp P}{\pp \drho_\mu} = \sum_{\nu} b^{-1}_{\mu \nu}( a_{\nu} P - J_{\nu})$
and obtain:
\begin{eqnarray}
\dot{S}^{\text{slow}}
&=&
\sum_{\mu}
\int_{\vecdrho} 
\dfrac{\pp J_\mu(\vecdrho,t)}{\pp \drho_{\mu}} 
\log\left[ P(\vecdrho,t) \right]
\nonumber \\
&=&
-
\sum_{\mu}
\int_{\vecdrho}
J_\mu(\vecdrho,t)
\dfrac{1}{P(\vecdrho,t)}  \dfrac{\pp P(\vecdrho,t)}{\pp \drho_{\mu}}
\nonumber \\
&=&
\Phi(t) + \Pi(t)
\mathcomma \label{eq:ent-prod-decomposition}
\end{eqnarray}
with
\begin{subequations}
\begin{eqnarray}
\Phi(t)
&=&
-
\sum_{\mu \nu}
\int_{\vecdrho}
J_\mu(\vecdrho,t) b_{\mu\nu}^{-1}(t) a_{\nu}(\vecdrho,t)\mathcomma
\nonumber \\ \label{eq:ent-flux}
\Pi(t)
&=&
\sum_{\mu \nu}
\int_{\vecdrho}
\dfrac{J_\mu(\vecdrho,t) b_{\mu\nu}^{-1}(t) J_{\nu}(\vecdrho,t)}{P(\vecdrho,t)}
\mathperiod \label{eq:ent-prod}
\end{eqnarray}
\end{subequations}
Thus, we can identify $\Pi(t) \geq 0$ as the entropy production rate of hydrodynamic modes.
$\Phi(t) \neq 0$ is caused by the entropy flux between the fast modes and the slow modes.

\section{\label{app:entr-prod-hyd-dim} Entropy production of hydrodynamic modes: dependence on dimensionality}

In this appendix, we examine the dependence of the hydrodynamic entropy production on the number of spatial dimensions $d$.
For simplicity, we consider a system with a single conserved mode ($\alpha(t)=0$)
and set the diffusion constant to the final value $D(t) \to D\fin \equiv 1$.
For small changes, the entropy production Eq.~\eqref{eq:ent-prod-tot} is
\begin{eqnarray}
\Pitot
&=&
\int_{-\infty}^{\infty} \! \dd t \, 
\int \! \dd^d q \, q^2 \, \dfrac{[C_q(t) -C^0(t)]^2}{C_q(t) C^0(t)}
\nonumber \\
&\approx &
\dfrac{1}{[C^{\text{fin}}]^2}
\int_{-\infty}^{\infty} \! \dd t \, 
\int \! \dd^d q \, q^2 [C_q(t) -C^0(t)]^2
\mathperiod \label{eq:entropy-prod-hyd-simple}
\end{eqnarray}
The convergence properties are easily analyzed in frequency space.
Using that $C_q(\omega) = \frac{2 q^2}{2 q^2 - \I \omega} C^0(\omega)$,
we obtain
\begin{eqnarray}
\Pitot 
& \propto &
\int_{-\infty}^{\infty} \! \dfrac{\dd \omega}{2\pi} 
\int \! \dd^d q \, |C^0(\omega)|^2 I_q(\omega)
\mathperiod \label{eq:entropy-prod-hyd-freq}
\end{eqnarray}
Following the terminology of \cite{Eckstein2010}
the behavior of $\Pitot$ depends on the ramp spectrum $|C^0(\omega)|^2$ and the intrinsic spectrum of hydrodynamic modes
\begin{eqnarray}
I_q(\omega)
&=&
\int \! \dd^d q \,
\dfrac{q^2\omega^2}{4q^4+\omega^2}
\propto
\int_0^{\Lambda} \! \dd q \,
\dfrac{ q^{d+1}\omega^2}{4q^4+\omega^2}
\nonumber \\
&\propto&
\begin{cases}
\omega^{3/2} & d=1 \\
\omega^2 \log\left( 1 + \dfrac{4 \Lambda^4}{\omega^2}\right) & d=2 \\
\omega^2 \Lambda^{d-2} & d>2
\end{cases}
\mathperiod \label{eq:entropy-prod-hyd-intrinsic}
\end{eqnarray}
The $q$ integral is UV-divergent for $d\geq 2$ and requires a cutoff $\Lambda$.
The $\omega$ integral is convergent for continuous ramps.
Even for ramps with non-analytic kinks (like the linear ramp in Eq.~\eqref{subeq:hopping}) $C^0(\omega)$ decays fast enough $\sim \omega^{-2}$ for large frequencies.
Thus, the behavior of $\Pitot(\tramp)$ does not depend on the shape of the ramp.
The scaling of the ramp spectrum $|\tilde{C}^0(\tilde{\omega})|^2 = \tramp^{-2} |C^0(\tilde{\omega} \tramp^{-1})|^2$
leads to 
the power-laws
\begin{eqnarray}
\Pitot(\tramp)
& \propto &
\begin{cases}
(\diffunit\tramp)^{-1/2} & d=1 \\
(\diffunit \tramp^{-1})\log\left( \Lambda^2 \diffunit\tramp \right)  & d=2 \\
(\diffunit\tramp)^{-1}  \Lambda^{d-2}& d>2
\end{cases}
\mathperiod \label{eq:entropy-prod-hyd-power-laws}
\end{eqnarray}
The hydrodynamic description of a quasi-adiabatic ramp breaks down for $d>2$ as fast modes dominate the entropy production.
Adding the full time-dependence in terms of the diffusion constant $D(t)$ and the source term $\propto \alpha(t)$ can change the prefactor by order $\landau(1)$, but does not affect the convergence properties of the integral \eqref{eq:entropy-prod-hyd-simple}.

\section{\label{app:entr-prod-Boltz} Entropy production of the Boltzmann equation for quasi-adiabatic parameter changes}

We contrast the hydrodynamic result with the prediction of the Boltzmann theory and calculate the change of the Boltzmann entropy of a classical system,
\begin{eqnarray}
S
&=&
-
\int \! \dd^d k \,
f_k \log\left( f_k \right)
\mathperiod \label{eq:entropy-prod-Boltz-entr}
\end{eqnarray}
We consider the particle-number- and energy-conserving relaxation time approximation
\begin{eqnarray}
\pd{t}f_k
&=& -\dfrac{f_k -f^0_k(t)}{\tsc}
\mathcomma \label{eq:entropy-prod-Boltz-RTA}
\end{eqnarray}
where $\tau$ is the relaxation time.
The equilibrium distribution function $f^0_k(t)= f^0[\dis{}{k}(t),\beta(t),z(t)]=\e^{-\beta(t)\dis{}{k}(t)}/z(t)$ depends on
the time-dependent dispersion $\dis{}{k}(t)$ (which follows a ramp protocol),
but also on the time-dependent inverse temperature $\beta(t)$ and the fugacity $z(t)$.
Particle-number and energy conservation is ensured by choosing $\beta(t)$ and $z(t)$ in such a way that
\begin{subequations} \label{subeq:entropy-prod-Boltz-sum-rule}
\begin{eqnarray}
\int \! \dd^d k \, \left[ f_k -f^0_k(t) \right]
&=& 0
\mathcomma \label{eq:entropy-prod-Boltz-sum-rule-N}
\\
\int \! \dd^d k \, \dis{}{k} \left[ f_k -f^0_k(t) \right]
&=& 0
\mathperiod \label{eq:entropy-prod-Boltz-sum-rule-E}
\end{eqnarray}
\end{subequations}
Using \eqref{eq:entropy-prod-Boltz-RTA} and \eqref{subeq:entropy-prod-Boltz-sum-rule}
we obtain the entropy production rate 
\begin{eqnarray}
\dot{S}
&=&
-
\left( \dfrac{L}{2 \pi} \right)^d 
\int \dd^d k \left( \pd{t} f_k \right) \log\left( f_k \right)
\nonumber \\
&=&
\left( \dfrac{L}{2 \pi} \right)^d \dfrac{1}{\tsc}
\int \! \dd^d k \, \delta f_k  \log\left( 1 + \dfrac{\delta f_k}{f^0_k(t)} \right)
\mathcomma \label{eq:entropy-prod-Boltz-rate}
\end{eqnarray}
with $\delta f_k = f_k -f^0_k(t)$. To derive \eqref{eq:entropy-prod-Boltz-rate},
we used that $\int \dd^d k \, \delta f_k  \log\left( f^0_k\right)=0$
which follows from $\log\left( f^0_k\right)=- \beta\dis{}{k} -\log(z)$ and \eqref{subeq:entropy-prod-Boltz-sum-rule}.
In the limit of a slow parameter change, $\tramp \gg \tau$, 
$\delta f_k \sim \frac{\tau}{\tramp}$ becomes smaller and smaller.
We can therefore expand in this parameter.
Terms linear in $\delta f_k$ vanish and thus $\pd{t} S \sim \frac{1}{\tramp^2}$.
Therefore, we recover $\Delta S=0$ in the adiabatic limit, $\tramp\to \infty$.
The leading contribution to the total entropy production is given by
\begin{eqnarray}
\Pitot
&\approx & \left( \dfrac{L}{2 \pi} \right)^d
\int_{-\infty}^{\infty} \! \dfrac{\dd t}{\tsc} \,
\int \! \dd^d k \,
\dfrac{\left(\delta f_k\right)^2}{f^0_k(t)}
\mathperiod \label{eq:entropy-prod-Boltz-total}
\end{eqnarray}
Using the Boltzmann equation \eqref{eq:entropy-prod-Boltz-RTA} and
$f_k\approx f^0_k$, we obtain $\delta f_k \approx -\tsc \pd{t} f^0_k(t)$.
A rescaling of the equilibrium distribution $\tilde{f}^0_k(\tilde{t})= f^0_k(\tilde{t} \,\tramp)$ with the dimensionless parameter $\tilde{t} = t/\tramp$ yields
\begin{eqnarray}
\Pitot
& \approx &
\frac{\tsc}{\tramp} \left( \dfrac{L}{2 \pi} \right)^d 
\int_{-\infty}^{\infty} \!\! \dd \tilde{t} \! 
\int \! \dd^d k \,\dfrac{[\pd{\tilde{t}}\tilde{f}^0_k(\tilde{t})]^2}{\tilde{f}^0_k(\tilde{t})}
\propto 
\frac{\tsc}{\tramp} 
\mathperiod \label{eq:entropy-prod-Boltz-total-rescaled}
\end{eqnarray}
To leading order in $1/\tramp$ one can use the adiabatic approximation, i.e, the assumption that the entropy remains constant in time, $S(t)=\text{const.}$, to calculate $\beta(t)$ and $z(t)$. Using these functions, one can directly calculate the entropy production using Eq.~\eqref{eq:entropy-prod-Boltz-total-rescaled}.
The integral is finite and does not diverge as long as $\dis{}{k}(t)$ is a continuous function.
Thus, we find that 
the total entropy production per volume is directly proportional to $\frac{\tsc}{\tramp}$.
Note that the Boltzmann equation misses all contributions from hydrodynamic fluctuations and therefore fails to predict the correct entropy production in $d\le 2$.

%
%
%
%
%
%
%

\section{\label{app:scaling} Scaling analysis}
To evaluate the importance of terms ignored within our analysis of the fluctuating diffusion equation, we perform
a scaling analysis with the goal to analyze the properties of the system in the limit of $\tramp\to \infty$. The starting point is the linear
diffusion equation, Eq.~\eqref{eq:linear-eq}.

We introduce dimensionless time coordinates $\tilde t=t/\tramp$ and also spatial coordinates, $\tilde \vecx=\vecx/\sqrt{\diffunit \tramp}$, where some diffusion constant $\diffunit$ is used for dimensional reasons and to match with the analysis given in Eq.~\eqref{subeq:ent-prod-tot-scaled}. 
A rescaled noise term $\tilde \veczeta_i = \veczeta_i \left(\frac{\tramp (\diffunit \tramp)^{d/2}}{\diffunit}\right)^{1/2}$ is chosen so that its correlation function is independent of $\tramp$ and depends only via the ratio $D_{ij}/\diffunit$ on diffusion constants. 
With this convention one obtains $\pdx \cdot \veczeta_i = \frac{1}{t_r (\diffunit \tramp)^{d/4}} \, \pdtildex \cdot \tilde \veczeta_i$, while $\pd{t} \drho_i=\frac{1}{\tramp} \pd{\tilde t} \drho_i$.
To obtain the same prefactor for these two terms, one has to rescale the densities according to $ \drhotilde_i=\drho_i  (\diffunit \tramp)^{d/4} $.
Now we use that $\alpha_{ij}(t)$ is proportional to the rate of change of the energy and therefore proportional to $1/\tramp$ to define $\tilde\alpha_{ij}(\tilde t)=\tramp\, \alpha_{ij}(\tramp\,\tilde t) $ which becomes independent of $\tramp$.
The correlation of the non-equilibrium noise $\eta_e$ defined in Eq.~\eqref{eq:noise-eta} are proportional to $1/\tramp^2$. Therefore
we introduce $\tilde \eta_e = \eta_e \, \tramp \left(\tramp (\diffunit \tramp)^{d/2}\right)^{1/2}$.
For $\tramp \to \infty$ the linearized diffusion equation Eq.~\eqref{eq:linear-eq} obtains in the new variables a form which is independent of $\tramp$:
\begin{align}
\pd{\tilde t} \drhotilde_i
&= \sum_j \frac{D_{ij}(\tilde t)}{\diffunit} \pdtildex^2 \drhotilde_j
- \sum_j \tilde \alpha_{ij}(\tilde t) \drhotilde_j
+ \pdtildex \cdot \tilde{\veczeta}_i
\mathperiod \label{eq:linear-eq-rescaled}
\end{align}
Note that we omitted the term proportional to the non-equilibrium noise $\tilde \eta$ as this obtains a prefactor $\tramp^{-1/2}$ and thus vanishes for $\tramp \to \infty$. In the language of renormalization group theory, 
this term is `irrelevant'. Importantly, one can use the scaling analysis to estimate the importance of terms omitted in the linearized diffusion equation, Eq.~\eqref{eq:linear-eq}. The higher-order derivative term $\pdx^4 \drho$,
for example, is suppressed by $1/(\diffunit \tramp)$.
The `leading irrelevant perturbations' and thus the most important correction arise from the density dependence of the diffusion constant and of the heating term proportional to $\pdx (\drho \pdx \drho)$ and $\frac{1}{\tramp} \drho^2$, respectively.
These terms give rise to corrections to Eq.~\eqref{eq:linear-eq-rescaled} proportional to
\begin{align}
\frac{1}{(\diffunit \tramp)^{d/4}} \pdtildex (\drhotilde \pdtildex \drhotilde) \qquad  \text{and} \qquad 
\frac{1}{(\diffunit \tramp)^{d/4}} \drhotilde^2 \mathperiod
\end{align}
Both scale to zero for $\tramp \to \infty$, but only very slowly, especially in $d=1$. To estimate the importance of such terms for the entropy production, it is important to note that the terms are even in $\drhotilde$ and the corresponding action is thus cubic in $\drhotilde$. Therefore,
they contribute only to quadratic order in perturbation theory (a fact not properly taken into account in Ref.~\cite{Lux2014}) and become equally important 
to corrections of the form $\frac{1}{(\diffunit \tramp)^{d/2}} \pdtildex (\drhotilde^2 \pdtildex \drhotilde)$ or   $ 
\frac{1}{(\diffunit \tramp)^{d/2}} \drhotilde^3$ which contribute to linear order. The correction to the entropy production from all these terms is suppressed by a factor $\frac{a^d}{(\diffunit \tramp)^{d/2}}$ relative to the leading-order term, where $a$ is some microscopic length scale, which cannot be obtained from our simple scaling analysis. As the leading correction in $d=1$ scales with $1/\sqrt{\diffunit t_r}$, the subleading term is proportional to $1/(\diffunit t_r)$ which motivates the higher-order fit used in \figref{fig:main-result}.

\section{\label{app:mom-con} Momentum conservation}

\begin{figure}[t]
\includegraphics{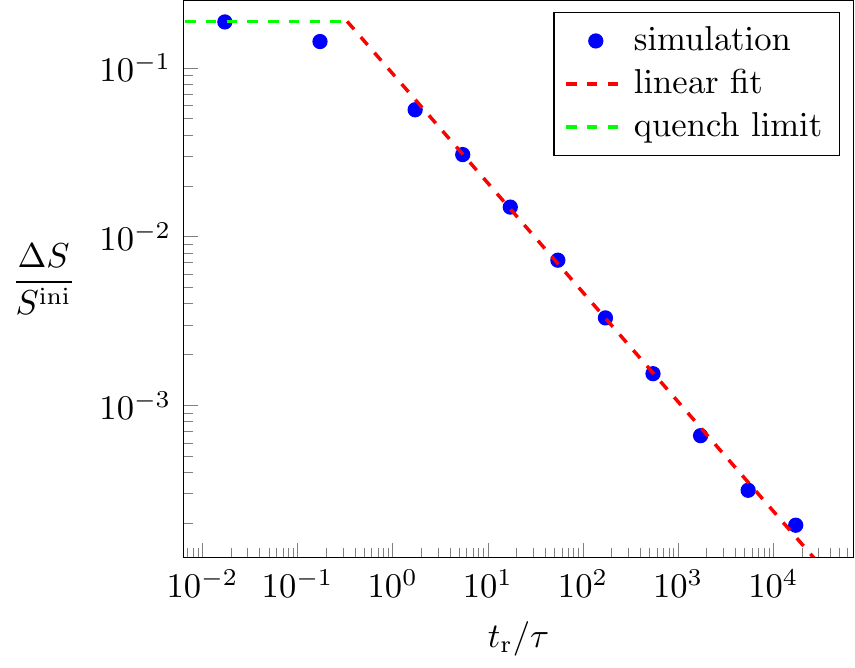}
\caption{\label{fig:result-mom-con} Similar to our main result \figref{fig:main-result}, the entropy production of the two-component gas decays algebraically for $\tramp \gg \taugamma$ if momentum conservation is added (blue dots).
However, the exponent changes: From a linear fit, we find $\Pitot \propto \tramp^{-0.65}$ (red-dashed line).
In the quench limit, the maximum entropy production of $\Pitot(\tramp \to 0)/S\ini \approx 0.19$ is reached (green-dashed line).
In our simulation, we used the time-dependent quadratic dispersion \eqref{eq:dis_quad_AB} and $\Gamma=0$.
}
\end{figure}

In the main text, we focused on the entropy production of diffusive systems.
Non-linear terms in the diffusion equation only contribute subleading corrections for these kind of systems as shown in \appref{app:scaling}.
In $d=1$ we found the leading-order power-law $\Pitot \propto \tramp^{-1/2}$.
The situation is different if one considers a system with momentum conservation. Here a new type of nonlinearity arises as the momentum current
(which can be identified with the pressure) obtains a correction proportional to $\drho^2$.
This term is relevant in $d=1$ and linear hydrodynamics breaks down \cite{Spohn1991,Praehofer2004,Mendl2013,Spohn2014}.
As a consequence, the exponent of the entropy production $\Pitot \propto \tramp^{-\plexp}$ is expected to differ from the value $\plexp=1/2$ obtained for diffusive systems.

To study the scaling of the entropy production in presence of momentum conservation,
we modify the dispersions of our two-component-gas model. 
We replace \eqref{eq:dis_AB} by a time-dependent quadratic dispersion,
\begin{eqnarray}
\dis{X}{k}(t)
&=&
J^{X}(t) \dfrac{k^2}{2} 
\mathcomma \label{eq:dis_quad_AB}
\end{eqnarray}
for particles of types $X=A,B$.
The momentum is conserved since umklapp scattering is absent. 
We still use the linear ramp protocol $J^{X}(t)$ defined by \eqref{subeq:hopping}.

\figref{fig:result-mom-con} shows the entropy production $\Pitot$ as function of $\tramp/\taugamma$ ($\Gamma=0$).
A fit $\Pitot \propto \tramp^{-\plexp}$ for large ramp times yields $\plexp \approx 0.65$. It is tempting to relate this exponent
to the exponents describing the decay of various correlation functions as function of time in thermal equilibrium which have been worked out by Spohn~\cite{Spohn2014}
for a momentum-conserving hydrodynamic theory.
Depending on which correlation function is considered, one obtains the exponents $2/3$, $5/3$ or $1/2$, typically with large subleading corrections. 
While our data is remarkably well described by the exponent $2/3$, further analytic insight is required to obtain a conclusive interpretation of the numerical data.

%
%
%
%

%
%
%


\bibliography{Adiabatic_quench_literature}

\begin{thebibliography}{37}%
\makeatletter
\providecommand \@ifxundefined [1]{%
 \@ifx{#1\undefined}
}%
\providecommand \@ifnum [1]{%
 \ifnum #1\expandafter \@firstoftwo
 \else \expandafter \@secondoftwo
 \fi
}%
\providecommand \@ifx [1]{%
 \ifx #1\expandafter \@firstoftwo
 \else \expandafter \@secondoftwo
 \fi
}%
\providecommand \natexlab [1]{#1}%
\providecommand \enquote  [1]{``#1''}%
\providecommand \bibnamefont  [1]{#1}%
\providecommand \bibfnamefont [1]{#1}%
\providecommand \citenamefont [1]{#1}%
\providecommand \href@noop [0]{\@secondoftwo}%
\providecommand \href [0]{\begingroup \@sanitize@url \@href}%
\providecommand \@href[1]{\@@startlink{#1}\@@href}%
\providecommand \@@href[1]{\endgroup#1\@@endlink}%
\providecommand \@sanitize@url [0]{\catcode `\\12\catcode `\$12\catcode
  `\&12\catcode `\#12\catcode `\^12\catcode `\_12\catcode `\%12\relax}%
\providecommand \@@startlink[1]{}%
\providecommand \@@endlink[0]{}%
\providecommand \url  [0]{\begingroup\@sanitize@url \@url }%
\providecommand \@url [1]{\endgroup\@href {#1}{\urlprefix }}%
\providecommand \urlprefix  [0]{URL }%
\providecommand \Eprint [0]{\href }%
\providecommand \doibase [0]{https://doi.org/}%
\providecommand \selectlanguage [0]{\@gobble}%
\providecommand \bibinfo  [0]{\@secondoftwo}%
\providecommand \bibfield  [0]{\@secondoftwo}%
\providecommand \translation [1]{[#1]}%
\providecommand \BibitemOpen [0]{}%
\providecommand \bibitemStop [0]{}%
\providecommand \bibitemNoStop [0]{.\EOS\space}%
\providecommand \EOS [0]{\spacefactor3000\relax}%
\providecommand \BibitemShut  [1]{\csname bibitem#1\endcsname}%
\let\auto@bib@innerbib\@empty
\bibitem [{\citenamefont {Born}\ and\ \citenamefont {Fock}(1928)}]{Born1928}%
  \BibitemOpen
  \bibfield  {author} {\bibinfo {author} {\bibfnamefont {M.}~\bibnamefont
  {Born}}\ and\ \bibinfo {author} {\bibfnamefont {V.}~\bibnamefont {Fock}},\
  }\bibfield  {title} {\bibinfo {title} {{Beweis des Adiabatensatzes}},\ }\href
  {https://doi.org/10.1007/BF01343193} {\bibfield  {journal} {\bibinfo
  {journal} {Zeitschrift f\"{u}r Physik}\ }\textbf {\bibinfo {volume} {51}},\
  \bibinfo {pages} {165} (\bibinfo {year} {1928})}\BibitemShut {NoStop}%
\bibitem [{\citenamefont {Kato}(1950)}]{Kato1950}%
  \BibitemOpen
  \bibfield  {author} {\bibinfo {author} {\bibfnamefont {T.}~\bibnamefont
  {Kato}},\ }\bibfield  {title} {\bibinfo {title} {{On the Adiabatic Theorem of
  Quantum Mechanics}},\ }\href {https://doi.org/10.1143/JPSJ.5.435} {\bibfield
  {journal} {\bibinfo  {journal} {J. Phys. Soc. Jpn.}\ }\textbf {\bibinfo
  {volume} {5}},\ \bibinfo {pages} {435} (\bibinfo {year} {1950})}\BibitemShut
  {NoStop}%
\bibitem [{\citenamefont {Avron}\ and\ \citenamefont
  {Elgart}(1999)}]{Avron1999}%
  \BibitemOpen
  \bibfield  {author} {\bibinfo {author} {\bibfnamefont {J.~E.}\ \bibnamefont
  {Avron}}\ and\ \bibinfo {author} {\bibfnamefont {A.}~\bibnamefont {Elgart}},\
  }\bibfield  {title} {\bibinfo {title} {{Adiabatic Theorem without a Gap
  Condition}},\ }\href {https://doi.org/10.1007/s002200050620} {\bibfield
  {journal} {\bibinfo  {journal} {Commun. Math. Phys.}\ }\textbf {\bibinfo
  {volume} {203}},\ \bibinfo {pages} {445} (\bibinfo {year}
  {1999})}\BibitemShut {NoStop}%
\bibitem [{\citenamefont {Callen}(1985)}]{Callen1985}%
  \BibitemOpen
  \bibfield  {author} {\bibinfo {author} {\bibfnamefont {H.~B.}\ \bibnamefont
  {Callen}},\ }\href {http://swbplus.bsz-bw.de/bsz012157619cov.htm} {\emph
  {\bibinfo {title} {Thermodynamics and an Introduction to
  Thermostatistics}}},\ \bibinfo {edition} {2nd}\ ed.\ (\bibinfo  {publisher}
  {Wiley},\ \bibinfo {address} {New York},\ \bibinfo {year} {1985})\BibitemShut
  {NoStop}%
\bibitem [{\citenamefont {Lieb}\ and\ \citenamefont
  {Yngvason}(1999)}]{Lieb1999}%
  \BibitemOpen
  \bibfield  {author} {\bibinfo {author} {\bibfnamefont {E.~H.}\ \bibnamefont
  {Lieb}}\ and\ \bibinfo {author} {\bibfnamefont {J.}~\bibnamefont
  {Yngvason}},\ }\bibfield  {title} {\bibinfo {title} {The physics and
  mathematics of the second law of thermodynamics},\ }\href
  {https://doi.org/https://doi.org/10.1016/S0370-1573(98)00082-9} {\bibfield
  {journal} {\bibinfo  {journal} {Phys. Rep.}\ }\textbf {\bibinfo {volume}
  {310}},\ \bibinfo {pages} {1} (\bibinfo {year} {1999})}\BibitemShut {NoStop}%
\bibitem [{\citenamefont {Bernier}\ \emph {et~al.}(2009)\citenamefont
  {Bernier}, \citenamefont {Kollath}, \citenamefont {Georges}, \citenamefont
  {De~Leo}, \citenamefont {Gerbier}, \citenamefont {Salomon},\ and\
  \citenamefont {K\"ohl}}]{Bernier2009}%
  \BibitemOpen
  \bibfield  {author} {\bibinfo {author} {\bibfnamefont {J.-S.}\ \bibnamefont
  {Bernier}}, \bibinfo {author} {\bibfnamefont {C.}~\bibnamefont {Kollath}},
  \bibinfo {author} {\bibfnamefont {A.}~\bibnamefont {Georges}}, \bibinfo
  {author} {\bibfnamefont {L.}~\bibnamefont {De~Leo}}, \bibinfo {author}
  {\bibfnamefont {F.}~\bibnamefont {Gerbier}}, \bibinfo {author} {\bibfnamefont
  {C.}~\bibnamefont {Salomon}},\ and\ \bibinfo {author} {\bibfnamefont
  {M.}~\bibnamefont {K\"ohl}},\ }\bibfield  {title} {\bibinfo {title} {Cooling
  fermionic atoms in optical lattices by shaping the confinement},\ }\href
  {https://doi.org/10.1103/PhysRevA.79.061601} {\bibfield  {journal} {\bibinfo
  {journal} {Phys. Rev. A}\ }\textbf {\bibinfo {volume} {79}},\ \bibinfo
  {pages} {061601} (\bibinfo {year} {2009})}\BibitemShut {NoStop}%
\bibitem [{\citenamefont {S\o{}rensen}\ \emph {et~al.}(2010)\citenamefont
  {S\o{}rensen}, \citenamefont {Altman}, \citenamefont {Gullans}, \citenamefont
  {Porto}, \citenamefont {Lukin},\ and\ \citenamefont {Demler}}]{Sorensen2010}%
  \BibitemOpen
  \bibfield  {author} {\bibinfo {author} {\bibfnamefont {A.~S.}\ \bibnamefont
  {S\o{}rensen}}, \bibinfo {author} {\bibfnamefont {E.}~\bibnamefont {Altman}},
  \bibinfo {author} {\bibfnamefont {M.}~\bibnamefont {Gullans}}, \bibinfo
  {author} {\bibfnamefont {J.~V.}\ \bibnamefont {Porto}}, \bibinfo {author}
  {\bibfnamefont {M.~D.}\ \bibnamefont {Lukin}},\ and\ \bibinfo {author}
  {\bibfnamefont {E.}~\bibnamefont {Demler}},\ }\bibfield  {title} {\bibinfo
  {title} {Adiabatic preparation of many-body states in optical lattices},\
  }\href {https://doi.org/10.1103/PhysRevA.81.061603} {\bibfield  {journal}
  {\bibinfo  {journal} {Phys. Rev. A}\ }\textbf {\bibinfo {volume} {81}},\
  \bibinfo {pages} {061603} (\bibinfo {year} {2010})}\BibitemShut {NoStop}%
\bibitem [{\citenamefont {Lubasch}\ \emph {et~al.}(2011)\citenamefont
  {Lubasch}, \citenamefont {Murg}, \citenamefont {Schneider}, \citenamefont
  {Cirac},\ and\ \citenamefont {Ba\~nuls}}]{Lubasch2011}%
  \BibitemOpen
  \bibfield  {author} {\bibinfo {author} {\bibfnamefont {M.}~\bibnamefont
  {Lubasch}}, \bibinfo {author} {\bibfnamefont {V.}~\bibnamefont {Murg}},
  \bibinfo {author} {\bibfnamefont {U.}~\bibnamefont {Schneider}}, \bibinfo
  {author} {\bibfnamefont {J.~I.}\ \bibnamefont {Cirac}},\ and\ \bibinfo
  {author} {\bibfnamefont {M.-C.}\ \bibnamefont {Ba\~nuls}},\ }\bibfield
  {title} {\bibinfo {title} {{Adiabatic Preparation of a Heisenberg
  Antiferromagnet Using an Optical Superlattice}},\ }\href
  {https://doi.org/10.1103/PhysRevLett.107.165301} {\bibfield  {journal}
  {\bibinfo  {journal} {Phys. Rev. Lett.}\ }\textbf {\bibinfo {volume} {107}},\
  \bibinfo {pages} {165301} (\bibinfo {year} {2011})}\BibitemShut {NoStop}%
\bibitem [{\citenamefont {Chiu}\ \emph {et~al.}(2018)\citenamefont {Chiu},
  \citenamefont {Ji}, \citenamefont {Mazurenko}, \citenamefont {Greif},\ and\
  \citenamefont {Greiner}}]{Chiu2018}%
  \BibitemOpen
  \bibfield  {author} {\bibinfo {author} {\bibfnamefont {C.~S.}\ \bibnamefont
  {Chiu}}, \bibinfo {author} {\bibfnamefont {G.}~\bibnamefont {Ji}}, \bibinfo
  {author} {\bibfnamefont {A.}~\bibnamefont {Mazurenko}}, \bibinfo {author}
  {\bibfnamefont {D.}~\bibnamefont {Greif}},\ and\ \bibinfo {author}
  {\bibfnamefont {M.}~\bibnamefont {Greiner}},\ }\bibfield  {title} {\bibinfo
  {title} {{Quantum State Engineering of a Hubbard System with Ultracold
  Fermions}},\ }\href {https://doi.org/10.1103/PhysRevLett.120.243201}
  {\bibfield  {journal} {\bibinfo  {journal} {Phys. Rev. Lett.}\ }\textbf
  {\bibinfo {volume} {120}},\ \bibinfo {pages} {243201} (\bibinfo {year}
  {2018})}\BibitemShut {NoStop}%
\bibitem [{\citenamefont {Landau}(1932)}]{Landau1932}%
  \BibitemOpen
  \bibfield  {author} {\bibinfo {author} {\bibfnamefont {L.~D.}\ \bibnamefont
  {Landau}},\ }\bibfield  {title} {\bibinfo {title} {On the theory of transfer
  of energy at collisions {II}},\ }\href@noop {} {\bibfield  {journal}
  {\bibinfo  {journal} {Phys. Z. Sowjetunion}\ }\textbf {\bibinfo {volume}
  {2}},\ \bibinfo {pages} {46} (\bibinfo {year} {1932})}\BibitemShut {NoStop}%
\bibitem [{\citenamefont {Zener}\ and\ \citenamefont
  {Fowler}(1932)}]{Zener1932}%
  \BibitemOpen
  \bibfield  {author} {\bibinfo {author} {\bibfnamefont {C.}~\bibnamefont
  {Zener}}\ and\ \bibinfo {author} {\bibfnamefont {R.~H.}\ \bibnamefont
  {Fowler}},\ }\bibfield  {title} {\bibinfo {title} {Non-adiabatic crossing of
  energy levels},\ }\href {https://doi.org/10.1098/rspa.1932.0165} {\bibfield
  {journal} {\bibinfo  {journal} {Proceedings of the Royal Society of London.
  Series A}\ }\textbf {\bibinfo {volume} {137}},\ \bibinfo {pages} {696}
  (\bibinfo {year} {1932})}\BibitemShut {NoStop}%
\bibitem [{\citenamefont {Altland}\ and\ \citenamefont
  {Gurarie}(2008)}]{Altland2008}%
  \BibitemOpen
  \bibfield  {author} {\bibinfo {author} {\bibfnamefont {A.}~\bibnamefont
  {Altland}}\ and\ \bibinfo {author} {\bibfnamefont {V.}~\bibnamefont
  {Gurarie}},\ }\bibfield  {title} {\bibinfo {title} {{Many Body Generalization
  of the Landau-Zener Problem}},\ }\href
  {https://doi.org/10.1103/PhysRevLett.100.063602} {\bibfield  {journal}
  {\bibinfo  {journal} {Phys. Rev. Lett.}\ }\textbf {\bibinfo {volume} {100}},\
  \bibinfo {pages} {063602} (\bibinfo {year} {2008})}\BibitemShut {NoStop}%
\bibitem [{\citenamefont {Polkovnikov}\ and\ \citenamefont
  {Gritsev}(2008)}]{Polkovnikov2008}%
  \BibitemOpen
  \bibfield  {author} {\bibinfo {author} {\bibfnamefont {A.}~\bibnamefont
  {Polkovnikov}}\ and\ \bibinfo {author} {\bibfnamefont {V.}~\bibnamefont
  {Gritsev}},\ }\bibfield  {title} {\bibinfo {title} {Breakdown of the
  adiabatic limit in low-dimensional gapless systems},\ }\href
  {https://doi.org/10.1038/nphys963} {\bibfield  {journal} {\bibinfo  {journal}
  {Nat. Phys.}\ }\textbf {\bibinfo {volume} {4}},\ \bibinfo {pages} {477}
  (\bibinfo {year} {2008})}\BibitemShut {NoStop}%
\bibitem [{\citenamefont {Moeckel}\ and\ \citenamefont
  {Kehrein}(2010)}]{Moeckel2010}%
  \BibitemOpen
  \bibfield  {author} {\bibinfo {author} {\bibfnamefont {M.}~\bibnamefont
  {Moeckel}}\ and\ \bibinfo {author} {\bibfnamefont {S.}~\bibnamefont
  {Kehrein}},\ }\bibfield  {title} {\bibinfo {title} {{Crossover from adiabatic
  to sudden interaction quenches in the Hubbard model: prethermalization and
  non-equilibrium dynamics}},\ }\href
  {https://doi.org/10.1088/1367-2630/12/5/055016} {\bibfield  {journal}
  {\bibinfo  {journal} {New J. Phys.}\ }\textbf {\bibinfo {volume} {12}},\
  \bibinfo {pages} {055016} (\bibinfo {year} {2010})}\BibitemShut {NoStop}%
\bibitem [{\citenamefont {Eckstein}\ and\ \citenamefont
  {Kollar}(2010)}]{Eckstein2010}%
  \BibitemOpen
  \bibfield  {author} {\bibinfo {author} {\bibfnamefont {M.}~\bibnamefont
  {Eckstein}}\ and\ \bibinfo {author} {\bibfnamefont {M.}~\bibnamefont
  {Kollar}},\ }\bibfield  {title} {\bibinfo {title} {Near-adiabatic parameter
  changes in correlated systems: influence of the ramp protocol on the
  excitation energy},\ }\href {https://doi.org/10.1088/1367-2630/12/5/055012}
  {\bibfield  {journal} {\bibinfo  {journal} {New J. Phys.}\ }\textbf {\bibinfo
  {volume} {12}},\ \bibinfo {pages} {055012} (\bibinfo {year}
  {2010})}\BibitemShut {NoStop}%
\bibitem [{\citenamefont {Dziarmaga}(2010)}]{Dziarmaga2010}%
  \BibitemOpen
  \bibfield  {author} {\bibinfo {author} {\bibfnamefont {J.}~\bibnamefont
  {Dziarmaga}},\ }\bibfield  {title} {\bibinfo {title} {Dynamics of a quantum
  phase transition and relaxation to a steady state},\ }\href
  {https://doi.org/10.1080/00018732.2010.514702} {\bibfield  {journal}
  {\bibinfo  {journal} {Adv. Phys.}\ }\textbf {\bibinfo {volume} {59}},\
  \bibinfo {pages} {1063} (\bibinfo {year} {2010})}\BibitemShut {NoStop}%
\bibitem [{\citenamefont {Deffner}(2017)}]{Deffner2017}%
  \BibitemOpen
  \bibfield  {author} {\bibinfo {author} {\bibfnamefont {S.}~\bibnamefont
  {Deffner}},\ }\bibfield  {title} {\bibinfo {title} {{Kibble-Zurek} scaling of
  the irreversible entropy production},\ }\href
  {https://doi.org/10.1103/PhysRevE.96.052125} {\bibfield  {journal} {\bibinfo
  {journal} {Phys. Rev. E}\ }\textbf {\bibinfo {volume} {96}},\ \bibinfo
  {pages} {052125} (\bibinfo {year} {2017})}\BibitemShut {NoStop}%
\bibitem [{\citenamefont {Alder}\ and\ \citenamefont
  {Wainwright}(1967)}]{Alder1967}%
  \BibitemOpen
  \bibfield  {author} {\bibinfo {author} {\bibfnamefont {B.~J.}\ \bibnamefont
  {Alder}}\ and\ \bibinfo {author} {\bibfnamefont {T.~E.}\ \bibnamefont
  {Wainwright}},\ }\bibfield  {title} {\bibinfo {title} {{Velocity
  Autocorrelations for Hard Spheres}},\ }\href
  {https://doi.org/10.1103/PhysRevLett.18.988} {\bibfield  {journal} {\bibinfo
  {journal} {Phys. Rev. Lett.}\ }\textbf {\bibinfo {volume} {18}},\ \bibinfo
  {pages} {988} (\bibinfo {year} {1967})}\BibitemShut {NoStop}%
\bibitem [{\citenamefont {Alder}\ and\ \citenamefont
  {Wainwright}(1970)}]{Alder1970}%
  \BibitemOpen
  \bibfield  {author} {\bibinfo {author} {\bibfnamefont {B.~J.}\ \bibnamefont
  {Alder}}\ and\ \bibinfo {author} {\bibfnamefont {T.~E.}\ \bibnamefont
  {Wainwright}},\ }\bibfield  {title} {\bibinfo {title} {{Decay of the Velocity
  Autocorrelation Function}},\ }\href {https://doi.org/10.1103/PhysRevA.1.18}
  {\bibfield  {journal} {\bibinfo  {journal} {Phys. Rev. A}\ }\textbf {\bibinfo
  {volume} {1}},\ \bibinfo {pages} {18} (\bibinfo {year} {1970})}\BibitemShut
  {NoStop}%
\bibitem [{\citenamefont {Ernst}\ \emph {et~al.}(1971)\citenamefont {Ernst},
  \citenamefont {Hauge},\ and\ \citenamefont {van Leeuwen}}]{Ernst1971}%
  \BibitemOpen
  \bibfield  {author} {\bibinfo {author} {\bibfnamefont {M.~H.}\ \bibnamefont
  {Ernst}}, \bibinfo {author} {\bibfnamefont {E.~H.}\ \bibnamefont {Hauge}},\
  and\ \bibinfo {author} {\bibfnamefont {J.~M.~J.}\ \bibnamefont {van
  Leeuwen}},\ }\bibfield  {title} {\bibinfo {title} {{Asymptotic Time Behavior
  of Correlation Functions. I. Kinetic Terms}},\ }\href
  {https://doi.org/10.1103/PhysRevA.4.2055} {\bibfield  {journal} {\bibinfo
  {journal} {Phys. Rev. A}\ }\textbf {\bibinfo {volume} {4}},\ \bibinfo {pages}
  {2055} (\bibinfo {year} {1971})}\BibitemShut {NoStop}%
\bibitem [{\citenamefont {Ernst}\ \emph
  {et~al.}(1976{\natexlab{a}})\citenamefont {Ernst}, \citenamefont {Hauge},\
  and\ \citenamefont {van Leeuwen}}]{Ernst1976}%
  \BibitemOpen
  \bibfield  {author} {\bibinfo {author} {\bibfnamefont {M.~H.}\ \bibnamefont
  {Ernst}}, \bibinfo {author} {\bibfnamefont {E.~H.}\ \bibnamefont {Hauge}},\
  and\ \bibinfo {author} {\bibfnamefont {J.~M.~J.}\ \bibnamefont {van
  Leeuwen}},\ }\bibfield  {title} {\bibinfo {title} {{Asymptotic time behavior
  of correlation functions. II. Kinetic and potential terms}},\ }\href
  {https://doi.org/10.1007/BF01012807} {\bibfield  {journal} {\bibinfo
  {journal} {J. Stat. Phys.}\ }\textbf {\bibinfo {volume} {15}},\ \bibinfo
  {pages} {7} (\bibinfo {year} {1976}{\natexlab{a}})}\BibitemShut {NoStop}%
\bibitem [{\citenamefont {Ernst}\ \emph
  {et~al.}(1976{\natexlab{b}})\citenamefont {Ernst}, \citenamefont {Hauge},\
  and\ \citenamefont {van Leeuwen}}]{Ernst1976a}%
  \BibitemOpen
  \bibfield  {author} {\bibinfo {author} {\bibfnamefont {M.~H.}\ \bibnamefont
  {Ernst}}, \bibinfo {author} {\bibfnamefont {E.~H.}\ \bibnamefont {Hauge}},\
  and\ \bibinfo {author} {\bibfnamefont {J.~M.~J.}\ \bibnamefont {van
  Leeuwen}},\ }\bibfield  {title} {\bibinfo {title} {{Asymptotic time behavior
  of correlation functions. III. Local equilibrium and mode-coupling theory}},\
  }\href {https://doi.org/10.1007/BF01012808} {\bibfield  {journal} {\bibinfo
  {journal} {J. Stat. Phys.}\ }\textbf {\bibinfo {volume} {15}},\ \bibinfo
  {pages} {23} (\bibinfo {year} {1976}{\natexlab{b}})}\BibitemShut {NoStop}%
\bibitem [{\citenamefont {Pomeau}\ and\ \citenamefont
  {R\'{e}sibois}(1975)}]{Pomeau1975}%
  \BibitemOpen
  \bibfield  {author} {\bibinfo {author} {\bibfnamefont {Y.}~\bibnamefont
  {Pomeau}}\ and\ \bibinfo {author} {\bibfnamefont {P.}~\bibnamefont
  {R\'{e}sibois}},\ }\bibfield  {title} {\bibinfo {title} {Time dependent
  correlation functions and mode-mode coupling theories},\ }\href
  {https://doi.org/https://doi.org/10.1016/0370-1573(75)90019-8} {\bibfield
  {journal} {\bibinfo  {journal} {Phys. Rep.}\ }\textbf {\bibinfo {volume}
  {19}},\ \bibinfo {pages} {63} (\bibinfo {year} {1975})}\BibitemShut {NoStop}%
\bibitem [{\citenamefont {Forster}(1975)}]{Forster1975}%
  \BibitemOpen
  \bibfield  {author} {\bibinfo {author} {\bibfnamefont {D.}~\bibnamefont
  {Forster}},\ }\href@noop {} {\emph {\bibinfo {title} {Hydrodynamic
  Fluctuations, Broken Symmetry, and Correlation Functions}}},\ Frontiers in
  physics ; 47\ (\bibinfo  {publisher} {Benjamin},\ \bibinfo {address}
  {Reading, Mass.},\ \bibinfo {year} {1975})\BibitemShut {NoStop}%
\bibitem [{\citenamefont {Harrison}\ and\ \citenamefont
  {Zwanzig}(1986)}]{Harrison1986}%
  \BibitemOpen
  \bibfield  {author} {\bibinfo {author} {\bibfnamefont {A.~K.}\ \bibnamefont
  {Harrison}}\ and\ \bibinfo {author} {\bibfnamefont {R.}~\bibnamefont
  {Zwanzig}},\ }\bibfield  {title} {\bibinfo {title} {On the relation between a
  fluctuating diffusion equation and long time tails in stationary random
  media},\ }\href {https://doi.org/10.1007/BF01010454} {\bibfield  {journal}
  {\bibinfo  {journal} {J. Stat. Phys.}\ }\textbf {\bibinfo {volume} {42}},\
  \bibinfo {pages} {935} (\bibinfo {year} {1986})}\BibitemShut {NoStop}%
\bibitem [{\citenamefont {Lux}\ \emph {et~al.}(2014)\citenamefont {Lux},
  \citenamefont {M\"uller}, \citenamefont {Mitra},\ and\ \citenamefont
  {Rosch}}]{Lux2014}%
  \BibitemOpen
  \bibfield  {author} {\bibinfo {author} {\bibfnamefont {J.}~\bibnamefont
  {Lux}}, \bibinfo {author} {\bibfnamefont {J.}~\bibnamefont {M\"uller}},
  \bibinfo {author} {\bibfnamefont {A.}~\bibnamefont {Mitra}},\ and\ \bibinfo
  {author} {\bibfnamefont {A.}~\bibnamefont {Rosch}},\ }\bibfield  {title}
  {\bibinfo {title} {Hydrodynamic long-time tails after a quantum quench},\
  }\href {https://doi.org/10.1103/PhysRevA.89.053608} {\bibfield  {journal}
  {\bibinfo  {journal} {Phys. Rev. A}\ }\textbf {\bibinfo {volume} {89}},\
  \bibinfo {pages} {053608} (\bibinfo {year} {2014})}\BibitemShut {NoStop}%
\bibitem [{\citenamefont {Spohn}(1991)}]{Spohn1991}%
  \BibitemOpen
  \bibfield  {author} {\bibinfo {author} {\bibfnamefont {H.}~\bibnamefont
  {Spohn}},\ }\href {http://zbmath.org/?q=an:0742.76002 ;
  http://swbplus.bsz-bw.de/bsz027050513cov.htm} {\emph {\bibinfo {title} {Large
  Scale Dynamics of Interacting Particles}}},\ Texts and Monographs in Physics\
  (\bibinfo  {publisher} {Springer},\ \bibinfo {address} {Berlin},\ \bibinfo
  {year} {1991})\BibitemShut {NoStop}%
\bibitem [{\citenamefont {Pr\"{a}hofer}\ and\ \citenamefont
  {Spohn}(2004)}]{Praehofer2004}%
  \BibitemOpen
  \bibfield  {author} {\bibinfo {author} {\bibfnamefont {M.}~\bibnamefont
  {Pr\"{a}hofer}}\ and\ \bibinfo {author} {\bibfnamefont {H.}~\bibnamefont
  {Spohn}},\ }\bibfield  {title} {\bibinfo {title} {{Exact Scaling Functions
  for One-Dimensional Stationary KPZ Growth}},\ }\href
  {https://doi.org/10.1023/B:JOSS.0000019810.21828.fc} {\bibfield  {journal}
  {\bibinfo  {journal} {J. Stat. Phys.}\ }\textbf {\bibinfo {volume} {115}},\
  \bibinfo {pages} {255} (\bibinfo {year} {2004})}\BibitemShut {NoStop}%
\bibitem [{\citenamefont {Mendl}\ and\ \citenamefont
  {Spohn}(2013)}]{Mendl2013}%
  \BibitemOpen
  \bibfield  {author} {\bibinfo {author} {\bibfnamefont {C.~B.}\ \bibnamefont
  {Mendl}}\ and\ \bibinfo {author} {\bibfnamefont {H.}~\bibnamefont {Spohn}},\
  }\bibfield  {title} {\bibinfo {title} {{Dynamic Correlators of
  Fermi-Pasta-Ulam Chains and Nonlinear Fluctuating Hydrodynamics}},\ }\href
  {https://doi.org/10.1103/PhysRevLett.111.230601} {\bibfield  {journal}
  {\bibinfo  {journal} {Phys. Rev. Lett.}\ }\textbf {\bibinfo {volume} {111}},\
  \bibinfo {pages} {230601} (\bibinfo {year} {2013})}\BibitemShut {NoStop}%
\bibitem [{\citenamefont {Spohn}(2014)}]{Spohn2014}%
  \BibitemOpen
  \bibfield  {author} {\bibinfo {author} {\bibfnamefont {H.}~\bibnamefont
  {Spohn}},\ }\bibfield  {title} {\bibinfo {title} {{Nonlinear Fluctuating
  Hydrodynamics for Anharmonic Chains}},\ }\href
  {https://doi.org/10.1007/s10955-014-0933-y} {\bibfield  {journal} {\bibinfo
  {journal} {J. Stat. Phys.}\ }\textbf {\bibinfo {volume} {154}},\ \bibinfo
  {pages} {1191} (\bibinfo {year} {2014})}\BibitemShut {NoStop}%
\bibitem [{\citenamefont {Prigogine}(1961)}]{Prigogine1967}%
  \BibitemOpen
  \bibfield  {author} {\bibinfo {author} {\bibfnamefont {I.}~\bibnamefont
  {Prigogine}},\ }\href@noop {} {\emph {\bibinfo {title} {Introduction to
  Thermodynamics of Irreversible Processes}}},\ \bibinfo {edition} {2nd}\ ed.\
  (\bibinfo  {publisher} {Interscience Publishers},\ \bibinfo {address} {New
  York},\ \bibinfo {year} {1961})\BibitemShut {NoStop}%
\bibitem [{\citenamefont {Tom\'{e}}(2006)}]{Tome2006}%
  \BibitemOpen
  \bibfield  {author} {\bibinfo {author} {\bibfnamefont {T.}~\bibnamefont
  {Tom\'{e}}},\ }\bibfield  {title} {\bibinfo {title} {{Entropy Production in
  Nonequilibrium Systems Described by a Fokker-Planck Equation}},\ }\href
  {http://www.scielo.br/scielo.php?script=sci_arttext&pid=S0103-97332006000700029&nrm=iso}
  {\bibfield  {journal} {\bibinfo  {journal} {Braz. J. Phys.}\ }\textbf
  {\bibinfo {volume} {36}},\ \bibinfo {pages} {1285} (\bibinfo {year}
  {2006})}\BibitemShut {NoStop}%
\bibitem [{\citenamefont {Jung}\ \emph {et~al.}(2006)\citenamefont {Jung},
  \citenamefont {Helmes},\ and\ \citenamefont {Rosch}}]{Jung2006}%
  \BibitemOpen
  \bibfield  {author} {\bibinfo {author} {\bibfnamefont {P.}~\bibnamefont
  {Jung}}, \bibinfo {author} {\bibfnamefont {R.~W.}\ \bibnamefont {Helmes}},\
  and\ \bibinfo {author} {\bibfnamefont {A.}~\bibnamefont {Rosch}},\ }\bibfield
   {title} {\bibinfo {title} {{Transport in Almost Integrable Models: Perturbed
  Heisenberg Chains}},\ }\href {https://doi.org/10.1103/PhysRevLett.96.067202}
  {\bibfield  {journal} {\bibinfo  {journal} {Phys. Rev. Lett.}\ }\textbf
  {\bibinfo {volume} {96}},\ \bibinfo {pages} {067202} (\bibinfo {year}
  {2006})}\BibitemShut {NoStop}%
\bibitem [{\citenamefont {Sachdev}\ and\ \citenamefont
  {Damle}(1997)}]{Sachdev1997}%
  \BibitemOpen
  \bibfield  {author} {\bibinfo {author} {\bibfnamefont {S.}~\bibnamefont
  {Sachdev}}\ and\ \bibinfo {author} {\bibfnamefont {K.}~\bibnamefont
  {Damle}},\ }\bibfield  {title} {\bibinfo {title} {{Low Temperature Spin
  Diffusion in the One-Dimensional Quantum $O(3)$ Nonlinear
  $\mathit{\ensuremath{\sigma}}$ Model}},\ }\href
  {https://doi.org/10.1103/PhysRevLett.78.943} {\bibfield  {journal} {\bibinfo
  {journal} {Phys. Rev. Lett.}\ }\textbf {\bibinfo {volume} {78}},\ \bibinfo
  {pages} {943} (\bibinfo {year} {1997})}\BibitemShut {NoStop}%
\bibitem [{\citenamefont {Damle}\ and\ \citenamefont
  {Sachdev}(1998)}]{Damle1998}%
  \BibitemOpen
  \bibfield  {author} {\bibinfo {author} {\bibfnamefont {K.}~\bibnamefont
  {Damle}}\ and\ \bibinfo {author} {\bibfnamefont {S.}~\bibnamefont
  {Sachdev}},\ }\bibfield  {title} {\bibinfo {title} {{Spin dynamics and
  transport in gapped one-dimensional Heisenberg antiferromagnets at nonzero
  temperatures}},\ }\href {https://doi.org/10.1103/PhysRevB.57.8307} {\bibfield
   {journal} {\bibinfo  {journal} {Phys. Rev. B}\ }\textbf {\bibinfo {volume}
  {57}},\ \bibinfo {pages} {8307} (\bibinfo {year} {1998})}\BibitemShut
  {NoStop}%
\bibitem [{\citenamefont {{Garst, M.}}\ and\ \citenamefont {{Rosch,
  A.}}(2001)}]{Garst2001}%
  \BibitemOpen
  \bibfield  {author} {\bibinfo {author} {\bibnamefont {{Garst, M.}}}\ and\
  \bibinfo {author} {\bibnamefont {{Rosch, A.}}},\ }\bibfield  {title}
  {\bibinfo {title} {{Transport in a classical model of a one-dimensional Mott
  insulator: Influence of conservation laws}},\ }\href
  {https://doi.org/10.1209/epl/i2001-00382-3} {\bibfield  {journal} {\bibinfo
  {journal} {Europhys. Lett.}\ }\textbf {\bibinfo {volume} {55}},\ \bibinfo
  {pages} {66} (\bibinfo {year} {2001})}\BibitemShut {NoStop}%
\bibitem [{\citenamefont {Rapp}\ \emph {et~al.}(2010)\citenamefont {Rapp},
  \citenamefont {Mandt},\ and\ \citenamefont {Rosch}}]{Rapp2010}%
  \BibitemOpen
  \bibfield  {author} {\bibinfo {author} {\bibfnamefont {A.}~\bibnamefont
  {Rapp}}, \bibinfo {author} {\bibfnamefont {S.}~\bibnamefont {Mandt}},\ and\
  \bibinfo {author} {\bibfnamefont {A.}~\bibnamefont {Rosch}},\ }\bibfield
  {title} {\bibinfo {title} {{Equilibration Rates and Negative Absolute
  Temperatures for Ultracold Atoms in Optical Lattices}},\ }\href
  {https://doi.org/10.1103/PhysRevLett.105.220405} {\bibfield  {journal}
  {\bibinfo  {journal} {Phys. Rev. Lett.}\ }\textbf {\bibinfo {volume} {105}},\
  \bibinfo {pages} {220405} (\bibinfo {year} {2010})}\BibitemShut {NoStop}%
\end{thebibliography}%

\end{document}